\documentclass[conference,10pt,letterpaper]{IEEEtran}

\IEEEoverridecommandlockouts

\usepackage{cite}
\usepackage{url}
\usepackage{listings}
\usepackage{caption}

\usepackage{amsmath,amssymb,amsfonts}
\usepackage{algorithmic}
\usepackage{graphicx}
\usepackage{textcomp}
\usepackage{xcolor}
\usepackage{booktabs,tabularx,array}
\newcolumntype{Y}{>{\raggedright\arraybackslash}X}
\usepackage[T1]{fontenc}
\usepackage[hidelinks]{hyperref}
\hypersetup{
    pdftitle={A Distributed View from a Single GPU:
    RankLens for Distributed Training},
    pdfauthor={Yufan Wang, Yihao Zhang, Yi Liu},
    pdfsubject={Single-Rank Emulation for Distributed Training},
    pdfkeywords={Distributed Training, Single-Rank Emulation,
    Communication Replay, Distributed Debugging}
}

\def\BibTeX{{\rm B\kern-.05em{\sc i\kern-.025em b}\kern-.08em
    T\kern-.1667em\lower.7ex\hbox{E}\kern-.125emX}}
\begin{document}

\bstctlcite{BSTcontrol}

\title{
A Distributed View from a Single GPU:\\
RankLens for Distributed Training%
\thanks{
This work has been submitted to the IEEE for possible
publication. Copyright may be transferred without notice,
after which this version may no longer be accessible.
}
}

\author{
\IEEEauthorblockN{Yufan Wang}
\IEEEauthorblockA{
\textit{Beihang University}\\
Beijing, China\\
yufanwang@buaa.edu.cn
}
\and
\IEEEauthorblockN{Yihao Zhang}
\IEEEauthorblockA{
\textit{Beihang University}\\
Beijing, China\\
sy2406433@buaa.edu.cn
}
\and
\IEEEauthorblockN{Yi Liu}
\IEEEauthorblockA{
\textit{Beihang University}\\
Beijing, China\\
yi.liu@buaa.edu.cn
}
}

\maketitle

\begin{abstract}
Modern distributed training is costly to debug, tune, and analyze because faithful analysis often requires repeated target-scale executions. We observe that many analysis tasks depend on the original distributed execution path, but not on full all-rank execution. We present RankLens, a single-rank emulation system that runs one survivor rank on a single GPU while preserving the target job’s distributed semantics. The survivor rank executes the original training program and performs real local computation. At communication boundaries, the system replaces real multi-rank communication with task-dependent semantics, selectively restoring recorded communication results and recorded or estimated completion timing while preserving the asynchronous dependencies required for analysis. This enables replay from real executions and supports analysis, debugging, and tuning with existing debuggers and performance analysis tools.

We evaluate the system on DeepSpeed and Megatron-LM across 8–64-GPU target configurations. RankLens achieves 100\% training-tensor communication-structure match, keeps emulation timing error below 5\%, supports local anomaly reproduction and execution analysis, and reduces the resource requirement of these analyses from target-scale clusters to a single GPU.
\end{abstract}


\section{Introduction}
The rapid scaling of AI models has placed increasing demands on training infrastructure, both in system complexity and resource requirements~\cite{grattafiori2024llama,deepseek2026v4,glm5_2026,kimi_k25_2026}. Modern distributed training systems combine data, tensor, and pipeline parallelism and use asynchronous communication to overlap computation with data transfer~\cite{narayanan2021efficient,li2020pytorch}. These mechanisms introduce complex data dependencies and synchronization constraints across ranks, making the numerical state and execution timing of each rank dependent on its peers~\cite{li2020pytorch}. Consequently, debugging and tuning tasks such as inspecting anomalous tensors, analyzing communication waits, and examining the overlap between computation and communication require the relevant distributed execution context, even when the analysis focuses on a single rank~\cite{pytorch_distributed_doc,hu2022dpro}. Reproducing and examining these behaviors through actual distributed execution typically requires provisioning the corresponding multi-GPU configuration, while repeated runs to collect additional observations or evaluate changes further increase resource costs. This creates a mismatch between the scope of analysis and its resource requirements: the behavior of interest may be local to a single rank, yet reproducing it remains tied to the full distributed configuration.

A straightforward way to reduce this cost is to simplify distributed training into conventional single-GPU execution. However, this changes the original parallel configuration and communication paths. To reduce reliance on full-cluster execution, prior work has developed performance models for estimating training costs~\cite{qi2017paleo,moolchandani2023amped,isaev2023calculon,geoffrey2021habitat,lee2025neusight}, as well as simulation and trace-based analysis tools for exploring training configurations, hardware designs, and optimizations~\cite{rashidi2020astra,won2023astra,zhu2020daydream,hu2022dpro,duan2024proteus,wang2025simai,li2025triosim,liang2025lumos,gui2025multiverse}. Recent systems further advance low-cost training analysis through execution modeling, framework code reuse, and hybrid emulation~\cite{purandare2025torchsim,qin2026phantora,xi2026prismllm}. These systems focus on reproducing target-scale execution behavior and interactions under limited resources, while rank-local debugging and tuning further require identifying which numerical states, timing effects, and execution dependencies must be preserved for direct inspection and evaluation of local changes. We therefore ask: how can the resources required for local analysis be reduced to a single GPU while preserving the numerical behavior, timing effects, and execution dependencies needed to reproduce, directly inspect, and tune the behavior of interest in the target rank's original code?

Our approach builds on a key observation: analyzing a target rank in distributed training does not necessarily require executing all ranks. The effects of remote ranks can be represented at communication boundaries by communication context, result values, and completion timing, while the target rank performs its actual local computation. Different analysis tasks require different subsets of these effects. Given the required communication effects and execution conditions, the target rank's original training code can therefore run on a single GPU, with the necessary communication effects restored to reproduce and analyze its local execution behavior.

Turning this observation into single-rank emulation for debugging and tuning is not as simple as bypassing peer-rank execution. The system must preserve the target rank's original distributed execution context, restore the required effects at the corresponding communication events, and preserve their asynchronous dependencies with subsequent computation~\cite{rajbhandari2020zero,narayanan2021efficient,li2020pytorch,pytorch_distributed_doc}. Distorting any of these properties can cause the replayed execution to diverge from the target execution, undermining its utility for debugging and tuning.

Building on this observation, we present \textbf{RankLens}, a single-rank emulation system for debugging and tuning distributed training. RankLens virtualizes distributed state to preserve the logical context of the target configuration, allowing the original training program to follow the target rank's distributed execution path on a single GPU while executing its local computation. At communication boundaries, RankLens maps collective and point-to-point (P2P) calls to virtual communication events, restores recorded communication results when required, and controls asynchronous completion through substituted \texttt{Work} objects. Communication timing can be restored from recorded executions or estimated from historical communication records when target-scale timing traces are unavailable or when tuning changes introduce communication works not covered by the original trace. The restored results are consumed by the original program, enabling developers to use existing debuggers (e.g., GDB~\cite{gdb_manual} and debugpy~\cite{debugpy}) to inspect anomalous tensors and their effects on subsequent computation and control flow. The same execution environment also supports analysis of computation-communication overlap and waiting behavior with Nsight Systems~\cite{nsight_systems}, while allowing developers to modify local computation or scheduling and directly observe the resulting timing changes.

Our main contributions are as follows:
\begin{itemize}
    \item \textbf{Single-rank distributed emulation.}
    We design and implement RankLens, which decouples the logical distributed configuration from physical GPU resources. RankLens preserves the target rank's logical identity, world size, process groups, and parallel configuration on a single GPU, allowing it to perform real local computation along the original distributed execution path.

    \item \textbf{An execution environment for local debugging, analysis, and tuning.}
    We retain real rank-local computation within the original distributed execution path, enabling existing debugging and profiling tools to inspect runtime states and execution behavior. Developers can also modify local computation or scheduling and directly observe the resulting timing changes, providing rapid feedback for tuning without repeatedly running the target-scale job.

    \item \textbf{Task-dependent communication semantic substitution.}
    We convert collective and P2P calls into virtual communication events and selectively restore their structural, numerical, and timing effects according to the analysis task, preserving the asynchronous execution dependencies required for analysis without communication with real peer ranks. Timing effects can be obtained from recorded executions or estimated from historical communication records when target-scale timing traces are unavailable.
\end{itemize}

\section{Background and Motivation}

\subsection{Distributed Training Execution and Communication Semantics}

Distributed training frameworks organize execution through ranks, process groups, parallelization strategies, and communication operations. For example, DeepSpeed ZeRO changes data layout and communication behavior by partitioning parameters, gradients, and optimizer states~\cite{rajbhandari2020zero}, while Megatron-LM combines multiple parallelism dimensions, such as data, tensor, and pipeline parallelism, with corresponding process groups~\cite{narayanan2021efficient}. Consequently, the distributed execution path is determined not only by the model architecture, but also by rank organization, process-group structure, communication order, synchronization points, and framework control logic.

Communication operations are embedded in this execution path and often exhibit asynchronous semantics. In PyTorch, for example, distributed communication may return an asynchronous \texttt{Work} whose result is consumed only at a later \texttt{wait} or synchronization point~\cite{li2020pytorch,pytorch_distributed_doc}. A communication operation therefore involves not only invocation, but also submission, completion, and synchronization, which jointly affect tensor states and the computation--communication timeline. Different analysis tasks depend on different communication semantics: communication-path analysis primarily relies on process-group context and invocation order, overlap analysis depends on communication completion and synchronization timing, and numerical debugging requires post-communication tensor states. Thus, low-resource debugging and tuning cannot simply reduce the number of GPUs; they must preserve the distributed execution context and communication semantics required by the target analysis.

\subsection{Motivating Debugging and Tuning Scenarios}

In large-scale distributed training, issues such as anomalous tensors, communication stalls, and insufficient computation--communication overlap often require further diagnosis using logs, profiling tools, or runtime checks. Obtaining finer-grained information typically requires repeatedly reproducing the relevant execution window with additional tracing or instrumentation~\cite{liu2026flightrecorder}. If each analysis iteration depends on the original distributed configuration, debugging and tuning remain tightly coupled to large-scale GPU resources.

Numerical debugging primarily depends on tensor states around communication boundaries. Training states such as gradients, parameter shards, and activations are propagated across ranks through collective or P2P communication~\cite{rajbhandari2020zero,narayanan2021efficient,li2020pytorch}, allowing numerical anomalies to spread across ranks. Further diagnosis requires associating an anomaly with the corresponding communication event and inspecting the communication result observed by the target rank and its effect on subsequent local computation. Such analysis therefore requires preserving both the communication context and the corresponding communication results.

Performance tuning requires evaluating how local optimizations affect the overall execution timeline. Operator replacement, fusion, or kernel optimization may reduce computation time while also changing computation--communication overlap, synchronization delays, and contention among GPU compute, memory, and communication resources~\cite{zhu2020daydream,hu2022dpro}. Such analysis therefore requires preserving the asynchronous dependencies among communication submission, completion, waiting, and local computation~\cite{li2020pytorch,pytorch_distributed_doc}. Moreover, tuning may change communication scheduling, message sizes, or invocation positions, leaving some communication \texttt{Work} objects without matching durations in the original trace; in such cases, their communication durations must be estimated.

\subsection{Scope and Requirements}

We focus on local debugging and tuning through single-rank emulation of a target distributed execution. Given a target rank, the goal is to reproduce and inspect the execution behavior relevant to a specific analysis task and evaluate local changes with limited resources, rather than reconstructing the complete all-rank execution. Therefore, the system only needs to preserve the effects of peer ranks that are visible to the target rank at communication boundaries and relevant to the analysis task.

Supporting such analysis imposes three requirements on single-rank replay. First, the target rank must continue along the original distributed execution path rather than falling back to conventional single-GPU training. Second, the system must restore the structure, value, or timing semantics required by the analysis task. Finally, these communication effects must be applied to the corresponding communication events while preserving the original asynchronous execution dependencies.

Accordingly, we scope the problem to replay-based emulation of a target distributed execution. RankLens restores structural and numerical effects from execution records to reproduce on a single GPU, the target rank's execution behavior relevant to the analysis task. Communication timing can either be replayed from recorded executions or estimated from historical communication records when target-scale timing traces are unavailable. The emulated execution assumes correspondence with the target execution in communication structure and control flow, allowing developers to inspect recorded behavior or evaluate local tuning changes within the preserved execution context.

\begin{table}[t]
\centering
\caption{Replay modes and their analysis scope. All modes execute the survivor rank's local computation.}
\label{tab:replay-modes}
\footnotesize
\begin{tabularx}{\columnwidth}{@{}>{\raggedright\arraybackslash}p{0.19\columnwidth}YY@{}}
\toprule
Mode & Required inputs & Supported analysis \\
\midrule
Structure & Execution context and communication metadata & Communication path and signature checks \\
Structure + timing & Above, plus recorded or estimated work durations & Step time, waits, and overlap \\
Value-enhanced & Above, plus selected recorded output payloads & Local numerical symptoms and their inspection \\
\bottomrule
\end{tabularx}
\end{table}

\section{System Design}

\subsection{System Overview}
\label{sec:system-overview}

\begin{figure*}[t]
    \centering
    \includegraphics[width=0.85\textwidth]{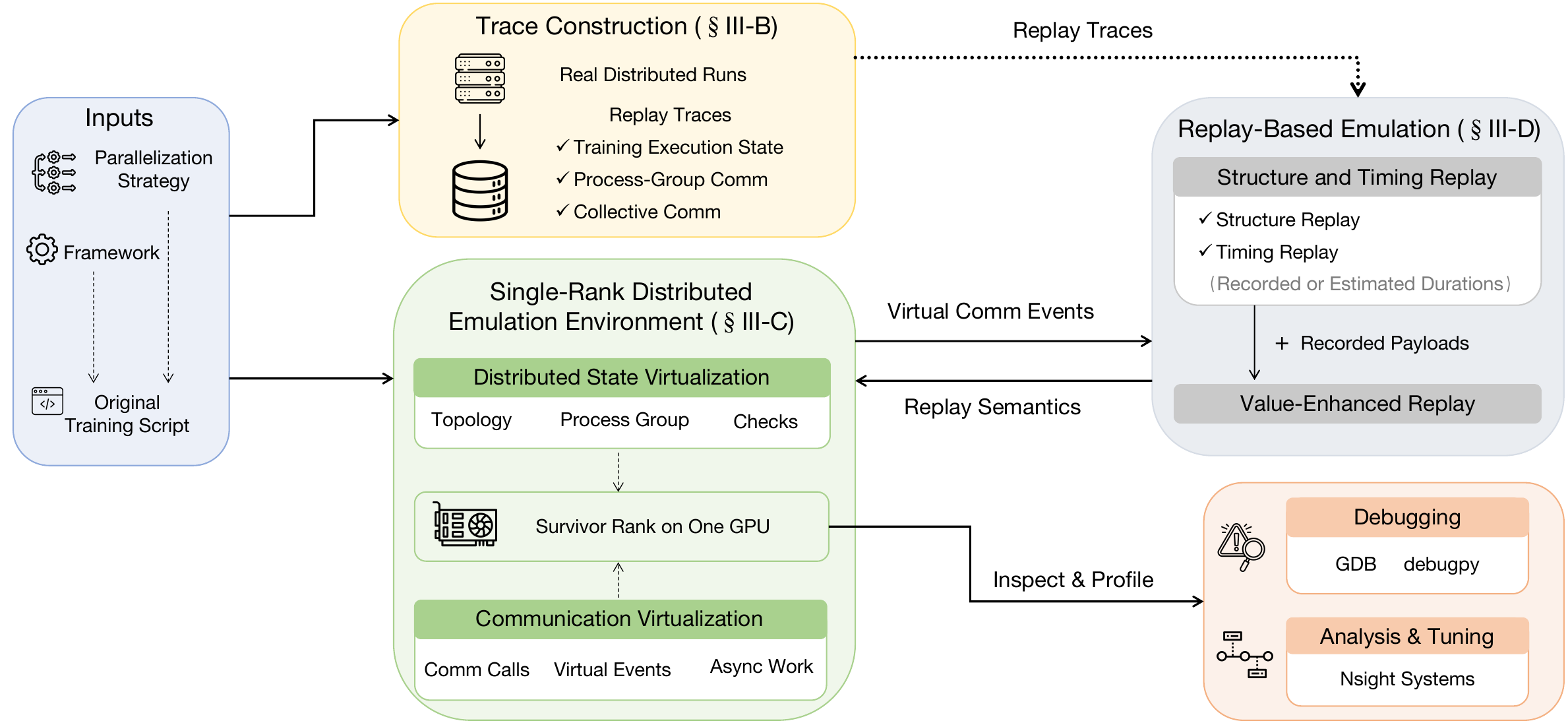}
    \caption{Overview of RankLens.}
    \label{fig:system-overview}
\end{figure*}

Figure~\ref{fig:system-overview} illustrates the overall architecture of \textbf{RankLens}, which performs emulation based on the original training script, distributed framework configuration, parallelization strategy, and target execution configuration. Depending on the analysis objective, the system constructs layered traces to provide the required communication effects (\S\ref{sec:trace-construction}).

After trace construction, the single-rank emulation engine runs the original user training script without modifying its model definition or training loop on a single GPU. Rather than falling back to ordinary single-GPU training, the survivor rank observes a virtualized distributed execution environment, allowing it to construct process groups according to the target configuration, enter the original distributed control flow, and trigger the corresponding collective and P2P operations and asynchronous communication work (\S\ref{sec:emulation-environment}). Within this environment, the survivor rank performs real local computation along the preserved execution path.

When the survivor rank issues a communication operation, the system replaces real multi-rank communication with task-dependent communication semantics at the communication boundary (\S\ref{sec:replay-emulation}). Structure and timing replay preserves communication structure and uses recorded or estimated communication timing to support performance analysis and tuning. Value-enhanced replay additionally restores recorded post-communication tensor states to support local debugging within the original program. Because local computation remains real execution, developers can modify local computation or scheduling and observe the resulting timing changes within the preserved distributed execution context. Through this single-rank execution with task-dependent communication substitution, the system converts a class of distributed training analysis tasks that would otherwise require large-scale clusters into executable emulation on a single GPU.

\subsection{Trace Construction}
\label{sec:trace-construction}

\begin{figure}[t]
\centering
\includegraphics[width=\columnwidth]{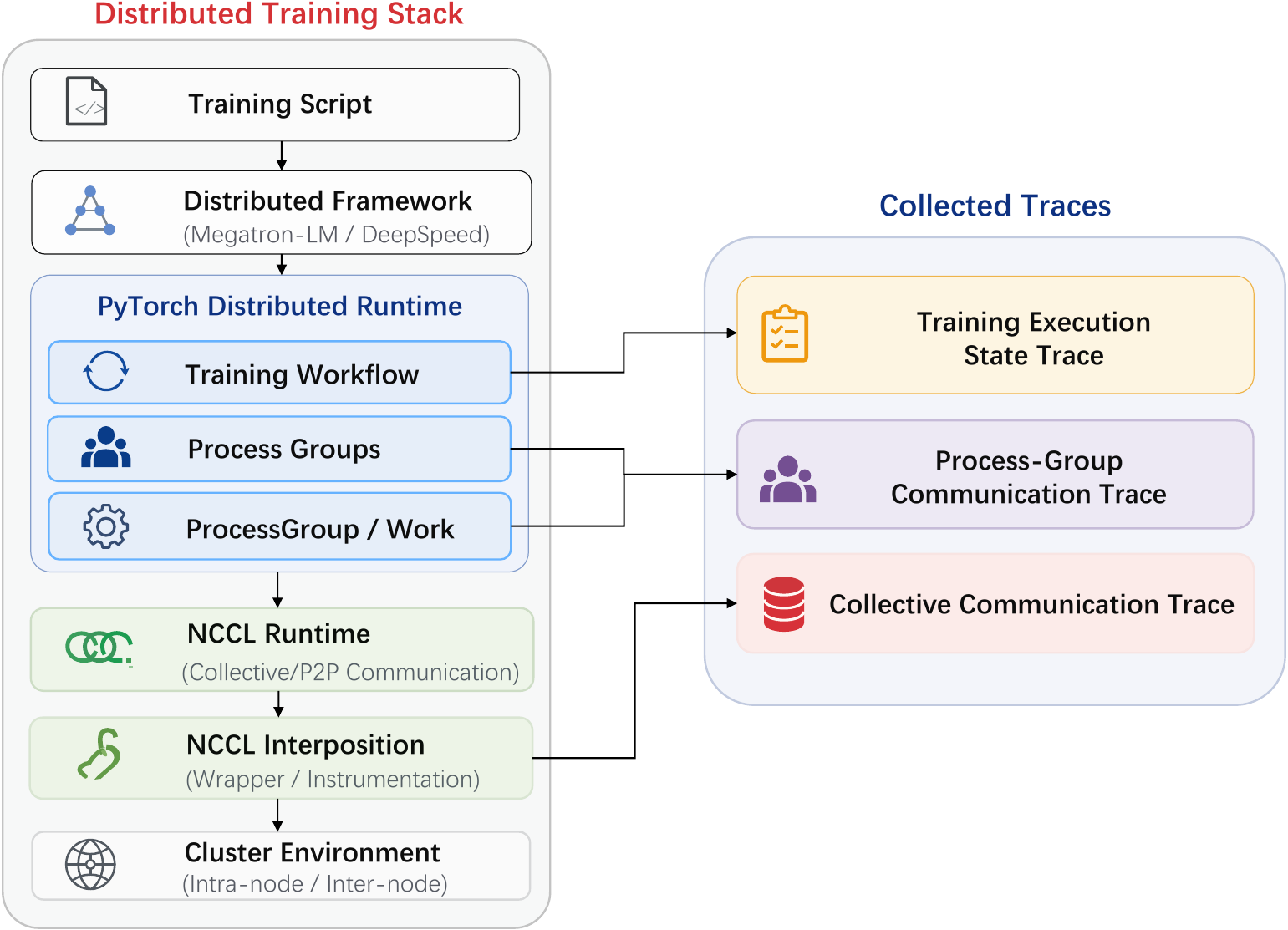}
\caption{Layered trace collection from real distributed training runs. Execution context and work metadata support event matching; work durations support timing replay; optional library-level payloads supply communication results for value-enhanced replay. Vertical arrows indicate the execution path, while horizontal arrows indicate trace collection at the corresponding layers.}
\label{fig:trace-construction}
\end{figure}

During real distributed training runs, the system collects three complementary traces: Training Execution State Trace, Process-Group Communication Trace, and Collective Communication Trace. They capture the training context, framework-visible communication works and their timing, and communication-library-level metadata and optional payloads, respectively. Together, they provide the information required for structure and timing replay and value-enhanced replay. Figure~\ref{fig:trace-construction} shows the layered trace collection workflow.

The \textbf{Training Execution State Trace} records the runtime context of a training job, including world size, global and local ranks, node placement, parallelization strategy, backend initialization state, and training-stage information such as epoch, step, forward pass, backward pass, gradient synchronization, and parameter update. This trace maps communication events to the original training execution flow.

The \textbf{Process-Group Communication Trace} records logical communication works visible to the distributed communication backend. For each collective or P2P work, it records the process-group context, rank mapping, operation type, message size, submission order, and measured duration. These durations characterize framework-visible communication works rather than individual low-level communication-library kernels. Together, the recorded metadata and durations provide the communication structure and timing information used for replay.

The \textbf{Collective Communication Trace} records collective-library calls and, when enabled, their buffer contents. It captures the communication interface, direction, buffer metadata, data size, and optional payload. This trace supports value-enhanced replay by providing recorded post-communication tensor states for local debugging.

\subsection{Single-Rank Distributed Emulation Environment}
\label{sec:emulation-environment}

Ordinary single-GPU training collapses framework-visible distributed state, changing process-group construction, communication calls, and synchronization behavior. To preserve the target distributed path with fewer resources, RankLens constructs a single-rank distributed emulation environment where one survivor rank runs on a single GPU while observing the target job's rank, world-size, topology, and process-group semantics. The environment combines \textbf{distributed state virtualization}, which preserves the logical state needed for initialization and distributed control flow, with \textbf{communication virtualization}, which turns collective and P2P operations into virtual communication events while preserving their interfaces, ordering, and asynchronous work lifecycle. Together, they keep the survivor rank on the target distributed path and expose communication boundaries for replay-based emulation (\S\ref{sec:replay-emulation}).

\subsubsection{Distributed State Virtualization}
\label{sec:state-virtualization}

Distributed state virtualization preserves the runtime semantics of the target distributed job for the survivor rank on a single GPU. Although physically executed on one GPU, the survivor rank is still exposed to the training framework as a logical rank in the target job. As a result, the framework follows the original initialization path, process-group construction, runtime checks, and distributed control flow, keeping the survivor rank on the original distributed execution path.

The system maintains two classes of virtualized state. Topology-level state represents the survivor rank's intra-node and inter-node placement in the target job, preserving its rank metadata and target topology metadata. Group-level state encodes the process-group semantics induced by the parallelization strategy, preserving the target process groups and their parallel-group metadata. The former defines rank placement and node boundaries, while the latter defines distributed component construction and collective and P2P call sites. Together, they provide the logical runtime state required for the survivor rank to follow the original distributed path.

While preserving the above logical state, the system decouples it from physical execution and GPU resources. The framework still observes the target world size, rank mapping, and process-group structure, but only the survivor rank executes the full training loop. The participation of non-survivor ranks in initialization, checks, and synchronization is replaced with deterministic local semantics, and framework-level logical device selection is mapped to the survivor GPU. This preserves the target distributed runtime semantics without provisioning the full set of training processes, GPU contexts, and model replicas required by the target job.

With these mechanisms, the survivor rank executes the framework-level distributed path as a logical rank in the target distributed job and reaches the corresponding communication boundaries. Distributed state virtualization preserves the distributed control flow up to these boundaries; communication virtualization then preserves the communication interfaces, ordering, and asynchronous lifecycle at the communication boundary.

\subsubsection{Communication Virtualization}
\label{sec:communication-virtualization}

Communication virtualization allows collective and P2P operations issued by the survivor rank to proceed without real multi-rank communication. When a communication call is issued along the original distributed control flow, RankLens intercepts it at the communication-library layer and maps it to a virtual communication event. The event records the communication context needed for subsequent emulation, including the communicator context, operation type, tensor metadata, message size, CUDA stream, and step-local sequence. This preserves the original communication invocation sites, signatures, and ordering while avoiding communication with peer ranks.

To preserve the communication-library interface, RankLens maintains lightweight virtual communicators that reflect the target rank and group context without being backed by physical multi-GPU resources. Intercepted calls preserve their communicator context, stream association, and grouped submission boundaries, but skip the actual multi-rank communication and return compatible local semantics. For value-sensitive analysis, the intercepted communication boundary also provides access to the corresponding buffers for optional restoration of recorded communication results. Timing and completion semantics are handled separately by the upper-level emulation backend.

\subsection{Replay-Based Emulation}
\label{sec:replay-emulation}

Replay-based emulation restores the communication effects required by single-rank execution without re-executing peer ranks. As described in \S\ref{sec:emulation-environment}, the emulation environment preserves the survivor rank's original distributed execution path and converts its collective and P2P calls into virtual communication events. According to the analysis objective, RankLens restores the required communication structure, timing, or values at the corresponding communication boundaries. Timing can be obtained from recorded executions or estimated from historical communication records, while value-enhanced replay restores recorded communication results at the communication-library layer.

The emulation engine first attempts to align each virtual communication event with an execution record using its training position, process-group context, operation signature, message size, and submission order. Matched records provide communication metadata, recorded durations, and optional payloads. For timing analysis, unmatched communication works can instead use durations estimated from historical communication records.

\subsubsection{Structure and Timing Replay}
\label{sec:structure-timing-replay}

In structure replay, execution remains driven by the original distributed training script. Dataset loading, sampler behavior, training loop execution, forward and backward computation, gradient synchronization, and parameter updates proceed through the original code, while the replay engine replaces each issued communication work with an emulated communication work. This preserves process-group semantics, communication invocation order, and per-step communication structure without restoring recorded payload contents. Structure replay uses communication metadata to align events with the recorded execution, but does not require timing injection. For configuration-related execution-path checks, the underlying emulation environment can also run the original script without a timing trace.

For timing analysis, each substituted Work object uses a matched recorded duration when available, or a duration estimated from historical traces otherwise. This permits preliminary timing analysis without a timing trace of the target execution, provided that communication substitution preserves the execution path relevant to the analysis. The engine controls when the work completes rather than blocking execution at communication submission, preserving the original submit, completion, and wait phases. The training program can therefore continue independent computation after communication submission, while dependent computation observes completion through the original synchronization interfaces. This extends structure replay with communication timing to support the analysis of communication waits and computation--communication overlap.

\subsubsection{Value-Enhanced Replay}
\label{sec:value-enhanced-replay}

For analyses that depend on post-communication tensor states, the system further supports value-enhanced replay. This extends structure and timing replay by additionally restoring recorded communication results. It uses the optional buffer contents in the Collective Communication Trace to restore post-communication tensors recorded in real runs. Rather than recomputing communication results from missing peer ranks, the system restores the recorded tensor states at the corresponding communication boundaries. In this work, value-enhanced replay uses durations and output payloads recorded from the same execution. Estimated durations are used only for timing analysis.

For each matched event with a recorded payload, the system restores the payload to the corresponding output buffer before subsequent computation consumes it. Value restoration must respect the asynchronous completion and synchronization dependencies established by the emulation environment, so that dependent computation observes the restored state. The original training program then continues execution using these values, while its local computation remains unchanged.

By combining recorded communication results with execution of the original program, value-enhanced replay supports the reproduction and inspection of local symptoms caused by recorded remote perturbations. Developers can use existing debuggers to set breakpoints, inspect post-communication tensors, and step through subsequent local computation to examine the effects of anomalous values. Such reproduction relies on the required local execution state and inputs, as well as correspondence between the replayed and recorded execution. It exposes the effects of recorded communication results on the survivor rank without reconstructing the execution of the remote ranks.

\subsection{Support for Existing Debugging and Performance Analysis Tools}
\label{sec:debugging-tuning}

RankLens preserves the target rank's original training code and real local computation, allowing existing debugging and performance analysis tools to be used directly. For numerical debugging, developers can use debugpy~\cite{debugpy} to inspect restored tensors and their effects on subsequent computation, while GDB~\cite{gdb_manual} can be used to examine native execution states and call paths in the underlying runtime. For performance analysis and tuning, Nsight Systems~\cite{nsight_systems} can be used to analyze communication waits, computation--communication overlap, and the impact of local code or scheduling changes on the execution timeline.

\section{Implementation}

RankLens realizes emulated execution at three layers: the framework runtime layer, the process-group communication backend layer, and the collective communication library layer.

At the framework runtime layer, RankLens applies lightweight patching to adapt selected runtime logic in distributed training frameworks, allowing the training program to preserve the logical state and execution flow required by the target distributed configuration on a single GPU. We make the necessary adaptations to PyTorch, DeepSpeed, and Megatron-LM, mainly around distributed initialization, state management, and framework control logic. As a result, the original training code can enter the single-rank emulation environment without modifying the model definition or main training loop.

At the process-group communication backend layer, RankLens implements a custom backend based on the PyTorch \texttt{ProcessGroup} interface. Its interface and \texttt{Work} lifecycle follow those of the native \texttt{ProcessGroupNCCL} backend, while introducing communication substitution and timing control for single-rank emulation. During real distributed execution, the backend records communication-work context, operation type, message size, and measured duration. During emulation, RankLens constructs a corresponding emulated communication \texttt{Work} for each communication call and injects the matched duration to control its logical completion time while preserving the original asynchronous dependencies.

When a measured duration is unavailable or a tuning change introduces an unmatched communication work, RankLens estimates its duration from historical records. It retrieves semantically similar samples according to features such as communication type, Process Group size, topology and rank locality, message size, and the relative position of the \texttt{Work} in the training execution flow. A local weighted regression is then applied to the neighboring samples to estimate the target communication duration, which is injected into the emulated \texttt{Work} through the same timing-control mechanism.

At the collective communication library layer, RankLens instruments NCCL communicator, group, collective, and P2P interfaces through dynamic function interception and symbol redirection, exposing low-level communication boundaries without modifying the upper-level training code. For each intercepted call, the system records information such as the operation type, communication context, buffer metadata, message size, and CUDA stream. During emulation, the actual multi-rank communication is bypassed and replaced with local semantics compatible with the original NCCL interface. For debugging tasks that require numerical semantics, RankLens can further restore recorded communication results to the corresponding buffers at these communication boundaries, allowing subsequent local computation to consume the restored states.

Listing~\ref{lst:nccl-wrapper} shows the simplified interception logic for \texttt{ncclAllReduce}.

\begin{lstlisting}[
    caption={Simplified wrapping logic for \texttt{ncclAllReduce}},
    label={lst:nccl-wrapper},
    captionpos=b,
    frame=single,
    framerule=0.4pt,
    framesep=6pt,
    basicstyle=\ttfamily\fontsize{7.5pt}{8.5pt}\selectfont,
    columns=fullflexible,
    keepspaces=true,
    showstringspaces=false,
    breaklines=false,
    xleftmargin=0pt,
    xrightmargin=0pt,
    aboveskip=6pt,
    belowskip=6pt
]
ncclAllReduce(sendbuff, recvbuff, count,
              datatype, op, comm, stream) {
    if (emulation_enabled) {
        if (value_replay_enabled) {
            restore_recorded_payload(
                recvbuff, stream);
        }
        return ncclSuccess;
    }

    record_nccl_metadata(
        comm, count, datatype, op, stream);

    status = original_func(
        sendbuff, recvbuff, count,
        datatype, op, comm, stream);

    if (status == ncclSuccess && capture_values) {
        capture_payload_after_completion(
            recvbuff, stream);
    }
    return status;
}
\end{lstlisting}

\section{Evaluation}
\label{sec:evaluation}

We evaluate single-rank emulation against real distributed executions, focusing on communication structure preservation, emulation timing accuracy, anomaly reproduction and local inspection, computation--communication overlap analysis, and trace collection overhead.

\subsection{Experimental Setup}
\label{sec:experimental-setup}

\begin{table}[t]
    \centering
    \caption{Testbed configuration.}
    \label{tab:testbed}
    \small
    \begin{tabular}{ll}
        \hline
        Hardware & Configuration \\
        \hline
        GPU & NVIDIA A100-PCIE-40GB \\
        Real-run scales & 2-64 GPUs \\
        Emulation hardware & 1 NVIDIA A100-PCIE-40GB \\
        Intra-node topology & PCIe-based, 4 GPUs per node \\
        Inter-node network & RoCEv2 \\
        NIC & Mellanox ConnectX-6 Dx, 100 Gb/s \\
        \hline
    \end{tabular}
\end{table}

\begin{table}[t]
    \centering
    \caption{Software stack.}
    \label{tab:software-stack}
    \small
    \begin{tabular}{ll}
        \hline
        Software & Version \\
        \hline
        CUDA & 12.4 \\
        cuDNN & 8.9.5 \\
        NCCL & 2.26.5 + CUDA 12.4 \\
        PyTorch & 2.7.0 \\
        PyTorch (controlled cases) & 2.7.0a0+git1341794 \\
        DeepSpeed & 0.18.9 \\
        Megatron-LM & 0.15.2 \\
        \hline
    \end{tabular}

    \vspace{2pt}
    \begin{minipage}{\columnwidth}
        \footnotesize
        The controlled cases use a locally compiled PyTorch
        build with the RankLens modifications and cuDNN 8.9.5.29.
    \end{minipage}
\end{table}

We conduct experiments on a multi-node A100 GPU cluster.
Each node is equipped with 4 NVIDIA A100-PCIE-40GB GPUs,
and nodes are connected through a RoCEv2 network.
Real distributed executions span 2-64 GPUs, while
single-rank replay runs on a single A100 GPU.
Table~\ref{tab:testbed} summarizes the testbed configuration.

Our experiments use DeepSpeed and Megatron-LM with
WikiText-2 and the GPT-2 BPE tokenizer.
Table~\ref{tab:software-stack} summarizes the software stack.
These model evaluations run the original distributed training
scripts without user-side modifications to model definitions
or training loops. The required runtime and communication-backend
adaptations are provided by the RankLens execution environment.
For each model and target configuration, we first perform
a real distributed execution to collect the traces required
for replay, and use it as ground truth.
We also use controlled training workloads to evaluate
anomaly reproduction, local inspection, and
computation-communication overlap analysis, with details
provided in the corresponding subsections.

\begin{figure}[t]
    \centering
    \includegraphics[width=\columnwidth]{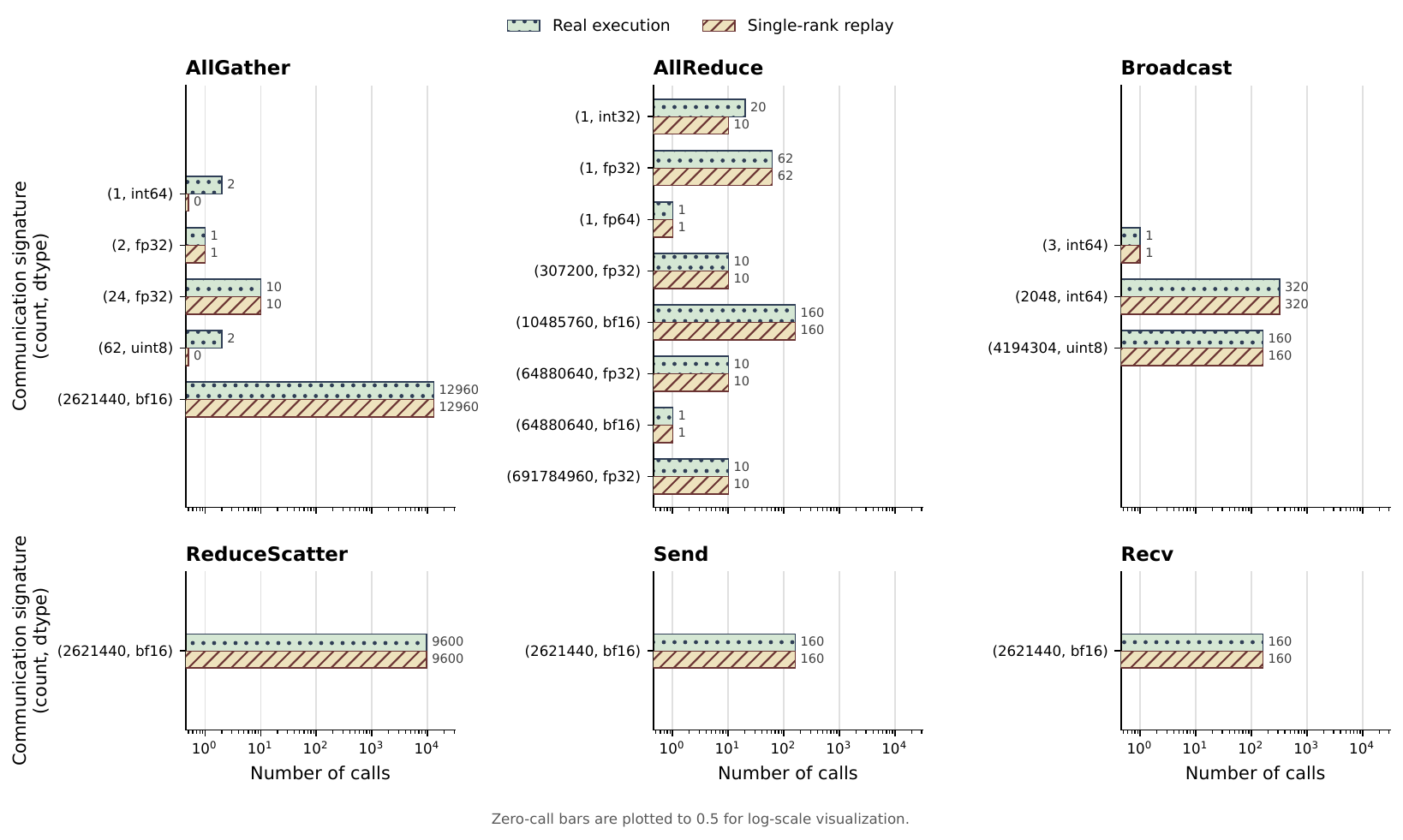}
    \caption{Raw signature comparison for Megatron-LM 10B on 64 GPUs with TP4/PP4.}
    \label{fig:signature_detail}
\end{figure}

\begin{figure*}[t]
    \centering
    \includegraphics[width=\textwidth]{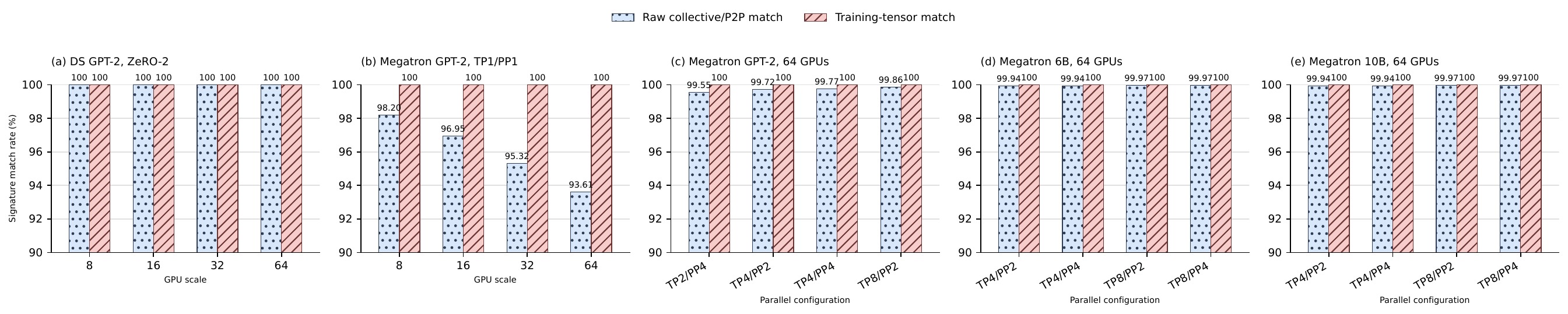}
    \caption{Signature match rates across DeepSpeed and Megatron-LM configurations.}
    \label{fig:signature_match}
\end{figure*}

\subsection{Communication Structure Preservation}
\label{sec:communication-structure}

We first evaluate whether single-rank replay preserves the communication structure observed in real distributed executions. This property is necessary because replayed durations and restored payloads are meaningful only when attached to the corresponding communication calls.

We represent each communication call with a communication signature:
\begin{equation}
    \sigma = (op, count, dtype, nbytes),
\end{equation}
where $op$ denotes the communication operation, $count$ is the element count, $dtype$ is the data type, and $nbytes$ is the payload size in bytes. We compare the number of occurrences of each signature in the real distributed execution and in single-rank replay. Runtime management APIs, such as \texttt{GET\_VERSION}, \texttt{GET\_UNIQUE\_ID}, \texttt{COMM\_INIT\_RANK}, and \texttt{COMM\_ABORT}, are excluded from the main training-communication metrics.

Figure~\ref{fig:signature_detail} shows the raw signature comparison for Megatron-LM 10B on 64 GPUs with TP4/PP4. The dominant training-tensor signatures match between real distributed execution and single-rank replay, including large \texttt{AllGather}, \texttt{ReduceScatter}, \texttt{AllReduce}, \texttt{Broadcast}, and pipeline \texttt{Send}/\texttt{Recv} calls. These matches cover sharded tensor communication, pipeline P2P messages, and full-precision synchronization. The few mismatches only appear in small metadata or control messages, such as the $(1,\mathrm{int64})$ \texttt{AllGather} signature. These calls do not carry model parameters, gradients, optimizer states, or pipeline activations and are excluded from the training-tensor metric below.

We summarize preservation across configurations using a call-weighted signature match rate:
\begin{equation}
    \operatorname{Match}(\Sigma) =
    \frac{
        \sum_{\sigma \in \Sigma}
        \min\left(N_{\mathrm{real}}(\sigma),
                  N_{\mathrm{replay}}(\sigma)\right)
    }{
        \sum_{\sigma \in \Sigma}
        \max\left(N_{\mathrm{real}}(\sigma),
                  N_{\mathrm{replay}}(\sigma)\right)
    },
\end{equation}
where $\Sigma$ is the set of signatures included in the comparison, and $N_{\mathrm{real}}(\sigma)$ and $N_{\mathrm{replay}}(\sigma)$ denote their occurrence counts in real execution and replay, respectively.

Figure~\ref{fig:signature_match} summarizes the signature match rates across all configurations. Raw collective/P2P match excludes runtime management APIs, while training-tensor match further excludes small metadata or control messages and counts only communication involving training tensors. DeepSpeed GPT-2 ZeRO-2 achieves 100\% for both metrics on 8--64 GPUs. Megatron GPT-2 TP1/PP1 has raw match rates between 93.61\% and 98.20\%, but its training-tensor match remains 100\%. For Megatron GPT-2 under 64-GPU TP/PP configurations, as well as the 6B and 10B models, raw match rates are close to 100\%, and training-tensor match remains 100\% in all cases. These results show that raw-level differences mainly come from non-training-tensor communication, while replay preserves the training-tensor signature counts across all evaluated configurations.

Overall, single-rank replay preserves the types and occurrence counts of training-tensor communication calls observed by the survivor rank, providing structural support for the following timing and behavioral evaluations.

\subsection{Emulation Timing Accuracy}
\label{sec:emulation-timing-accuracy}

\begin{figure}[t]
  \centering
  \includegraphics[width=\columnwidth]{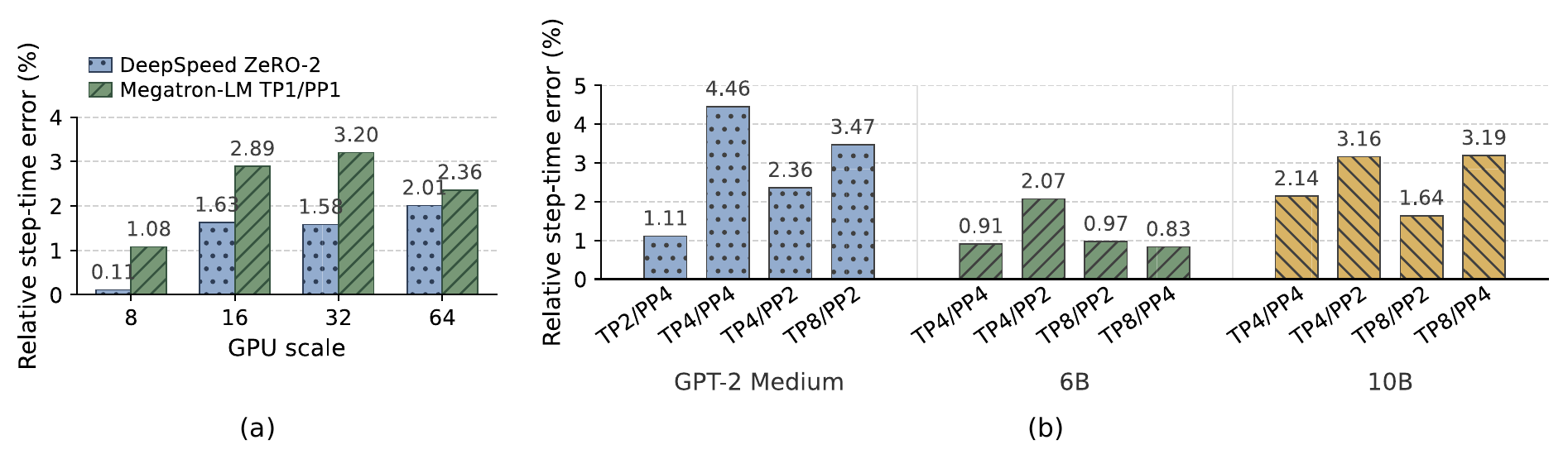}
  \caption{Emulation timing accuracy with recorded durations:
  (a) across GPU scales; (b) under 64-GPU Megatron-LM configurations.}
  \label{fig:timing_combined}
\end{figure}

\begin{figure}[t]
  \centering
  \includegraphics[width=\columnwidth]{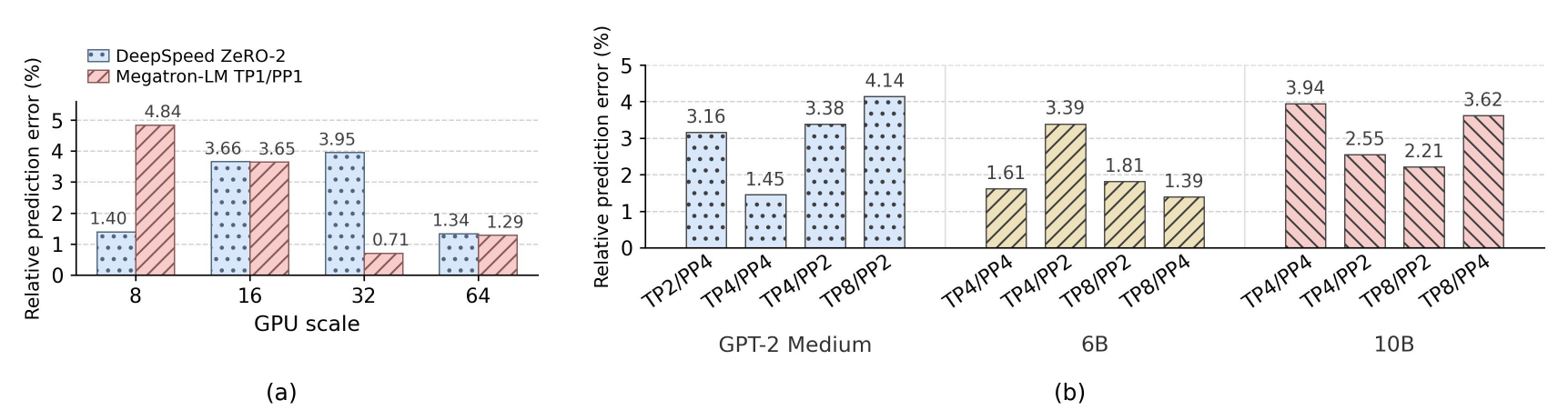}
  \caption{Emulation timing accuracy with estimated communication durations: (a) across GPU scales; (b) under 64-GPU Megatron-LM configurations.}
  \label{fig:estimated_timing_combined}
\end{figure}

After evaluating communication-structure preservation, we further evaluate the timing accuracy of RankLens. Timing emulation uses recorded durations when available and estimated durations for unmatched communication works. Both are injected through the same emulated communication \texttt{Work} mechanism. We use the relative step-time error:
\begin{equation}
    \operatorname{Error} =
    \frac{\left|T_{\mathrm{emu}} - T_{\mathrm{real}}\right|}
         {T_{\mathrm{real}}},
\end{equation}
where $T_{\mathrm{real}}$ is the step time of the real distributed execution and $T_{\mathrm{emu}}$ is the step time of single-rank emulation.

With recorded durations, Figure~\ref{fig:timing_combined}(a) shows errors of 0.11\%--2.01\% for DeepSpeed ZeRO-2 and 1.08\%--3.20\% for Megatron-LM TP1/PP1 across 8--64 GPUs. Figure~\ref{fig:timing_combined}(b) further shows that the error remains below 4.46\% across different 64-GPU Megatron-LM model sizes and TP/PP configurations.

With estimated durations, Figure~\ref{fig:estimated_timing_combined}(a) shows errors of 1.34\%--3.95\% for DeepSpeed ZeRO-2 and 0.71\%--4.84\% for Megatron-LM TP1/PP1. Figure~\ref{fig:estimated_timing_combined}(b) shows errors of 1.39\%--4.14\% across the evaluated 64-GPU Megatron-LM configurations. These results show that both recorded and estimated communication durations provide sufficiently accurate timing for single-rank emulation. Estimated durations therefore extend timing-aware emulation to communication works not covered by the original trace.

\subsection{Anomaly Reproduction and Local Inspection}
\label{sec:anomaly-reproduction}

\begin{table}[t]
    \centering
    \caption{Reproduction of a gradient anomaly and its effect
    on the original SGD update.}
    \label{tab:anomaly-reproduction}
    \begingroup
    \footnotesize
    \setlength{\tabcolsep}{3pt}
    \renewcommand{\arraystretch}{1.15}
    \begin{tabularx}{\columnwidth}{
        @{}
        >{\raggedright\arraybackslash}X
        cc
        >{\raggedright\arraybackslash}p{0.22\columnwidth}
        @{}
    }
        \toprule
        Execution
        & \shortstack{Non-finite\\gradients}
        & \shortstack{Non-finite\\parameters\\after SGD}
        & \shortstack[l]{Agreement\\with reference} \\
        \midrule
        \shortstack[l]{Real /\\Normal}
        & 0 & 0 & Reference \\

        \shortstack[l]{Real /\\Fault injected}
        & 1 & 1 & Reference \\

        \shortstack[l]{Replay /\\Faulty payload}
        & 1 & 1 & Match$^{*}$ \\

        \shortstack[l]{Replay /\\Normal payload}
        & 0 & 0 & Exact match \\
        \bottomrule
    \end{tabularx}

    \par\smallskip
    \begin{minipage}{\columnwidth}
        \footnotesize
        Each condition is repeated across three independent runs.
        Comparisons cover all 577 gradient and parameter elements.
        $^{*}$NaN positions match the fault-injected reference,
        and the maximum absolute error over finite elements is zero.
        Normal replay exactly matches the normal reference.
    \end{minipage}
    \endgroup
\end{table}

We use a controlled DDP training case to evaluate whether RankLens can restore anomalous communication results and reproduce their effects on subsequent local computation. The workload contains 577 FP32 parameters and uses SGD for parameter updates. During real two-GPU execution, we set the first element of rank~1's gradient bucket to NaN and record the result received by rank~0 after AllReduce. RankLens then restores this result on a single GPU, where the original program performs the SGD update. We compare normal and fault-injected real executions with single-GPU replay using the corresponding normal and faulty payloads. Each condition uses the same initialization and inputs and is repeated across three independent runs.

Table~\ref{tab:anomaly-reproduction} shows that the anomalous gradient in the fault-injected real execution causes one parameter to become NaN after the update. Replay with the faulty payload reproduces the same NaN positions in the gradients and updated parameters, while all remaining finite elements match the fault-injected execution exactly. Replay with the normal payload exactly matches the normal real execution, with no non-finite values. The comparisons cover all 577 gradient and parameter elements, with consistent results across the three runs. These results demonstrate that RankLens reproduces both the anomalous communication result and its effect on the original SGD update.

During replay, \texttt{debugpy} exposes the anomalous gradient before the SGD update and the corresponding anomalous parameter afterward. GDB reaches the wrapper entry for the target AllReduce and the internal payload-restoration function, exposing the call path from \texttt{ProcessGroupNCCL} to the communication-substitution layer. Developers can thus inspect the restored tensors and their effects on local behavior on a single GPU. The remote fault location is specified by the injection procedure; this experiment does not evaluate automatic remote root-cause localization.

\subsection{Computation-Communication Overlap Analysis and Tuning}
\label{sec:overlap-analysis}

\begin{figure}[t]
    \centering
    \includegraphics[width=\columnwidth]{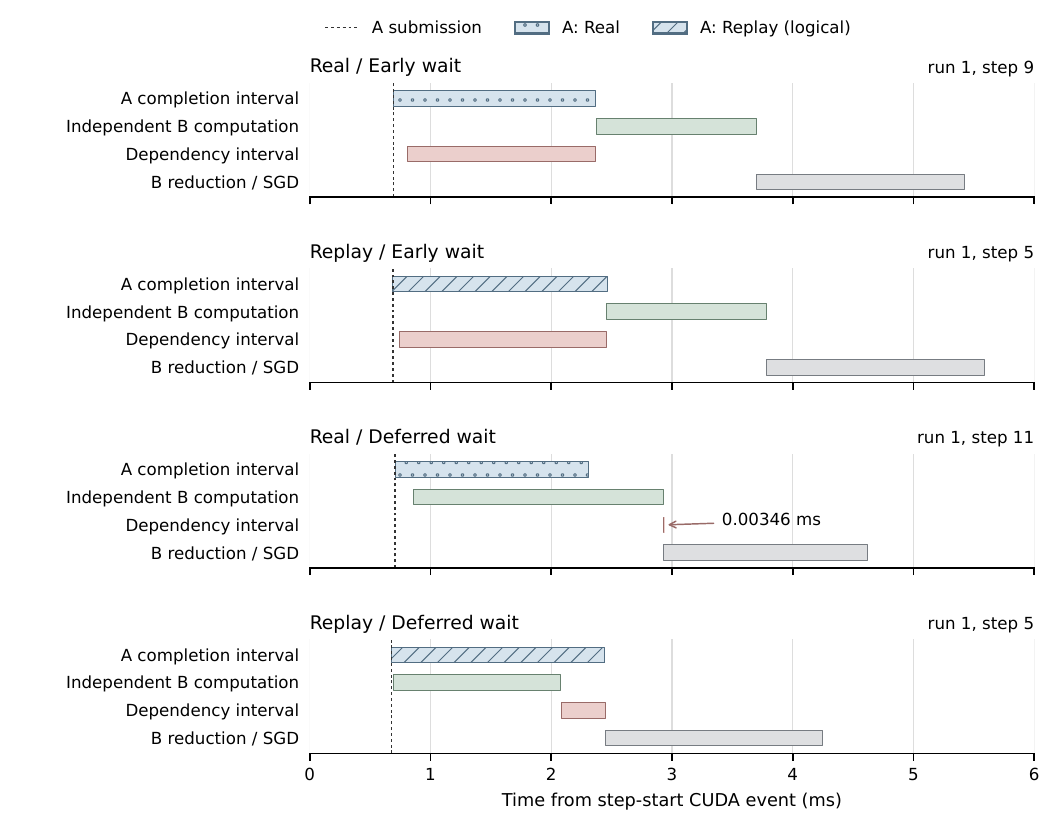}
    \caption{Representative step timelines under early and
    deferred waiting, shown on a shared time scale.
    Deferring the wait allows independent computation in
    branch B to overlap with branch A's outstanding communication.
    Blue intervals span submission to observed completion
    in real execution; hatched intervals denote logical
    completion timing in replay, not NCCL kernel execution.
    Red intervals are delimited by CUDA events around the
    wait and may include submission gaps.
    Gray intervals cover branch B's reduction and the SGD update.}
    \label{fig:overlap-timeline}
\end{figure}

\begin{figure}[t]
    \centering
    \includegraphics[width=\columnwidth]{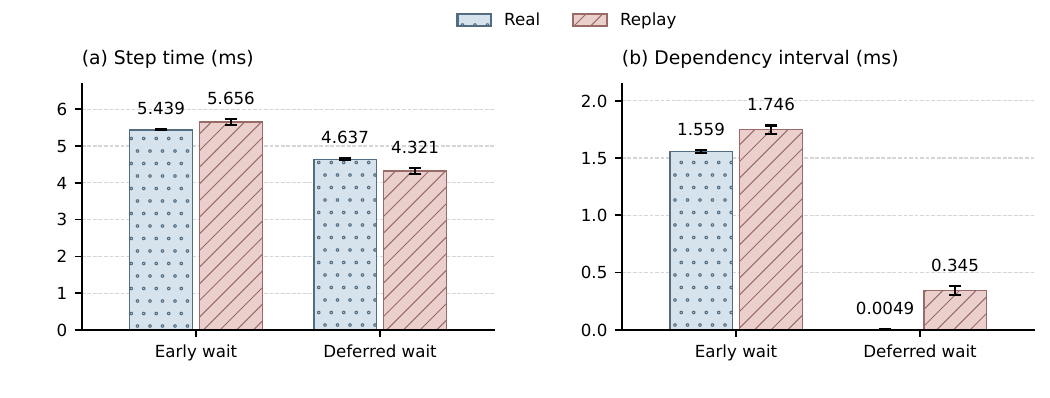}
    \caption{Effect of deferring communication waits on
    (a) step time and (b) the dependency interval.
    Bars show the mean across five independent runs;
    error bars indicate the standard deviation of the per-run
    means, each computed over twenty measured steps.
    Both replay schedules use the same trace collected
    under the early-wait schedule.}
    \label{fig:overlap-results}
\end{figure}

We use a controlled two-branch training workload to evaluate RankLens's support for overlap analysis and tuning. Computation in branch B does not depend on the result of branch A's gradient reduction, allowing the wait for A's communication to be deferred from before B's computation to afterward, before the parameter update. Both single-GPU replay schedules use the same trace collected from real execution with early waiting, and real two-GPU runs validate the scheduling change. Each condition is evaluated across five independent runs, each with five warm-up steps followed by twenty measured steps. We report the mean and standard deviation of the per-run means.

Figure~\ref{fig:overlap-timeline} shows that deferring the wait allows B's independent computation to overlap with A's outstanding communication, shortening the dependency interval. This interval is delimited by CUDA events placed before and after the wait and may include submission gaps. In replay, the communication interval represents logical completion timing rather than actual NCCL kernel execution.

Figure~\ref{fig:overlap-results} shows that deferring the wait reduces replay step time from 5.656 ms to 4.321 ms and real-execution step time from 5.439 ms to 4.637 ms. The corresponding reductions, computed from paired runs, are 23.60\% and 14.75\%, respectively. Replay correctly identifies the direction of improvement but overestimates its magnitude by 8.85 percentage points. The computation interval of branch B increases more in real execution, consistent with resource contention between computation and communication that replay does not fully capture. These results demonstrate that RankLens can support computation-communication overlap analysis and help identify beneficial scheduling changes.

\subsection{Trace Collection Overhead}
\label{sec:trace-collection-overhead}

\begin{figure}[t]
  \centering
  \includegraphics[width=\columnwidth]{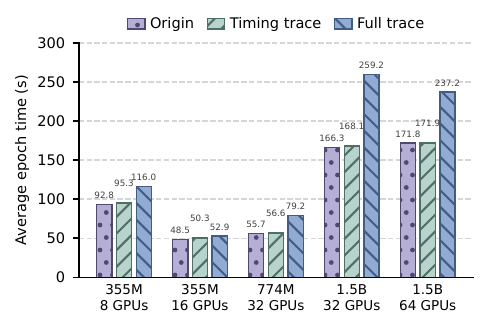}
  \caption{Training-time impact of trace collection. The figure compares the uninstrumented baseline, timing trace, and full trace. Timing trace records execution context, communication metadata, and timing without tensor payloads; full trace additionally captures tensor payloads using asynchronous I/O.}
  \label{fig:full_trace_collection_overhead}
\end{figure}

Figure~\ref{fig:full_trace_collection_overhead} reports the average epoch time under the three trace collection modes. Timing trace introduces little overhead across all evaluated configurations: for 355M-1.5B models on 8-64 GPUs, its overhead over the origin run remains below 4\%. This shows that the default tracing path, which records execution context, communication metadata, and timing without tensor payloads, does not materially perturb the real distributed executions used as ground truth, and is sufficient for structure and timing replay.

In contrast, full trace incurs higher overhead because it additionally captures tensor payloads. The overhead increases with model scale and communication volume, but remains bounded overall. Since payloads are only required by value-enhanced replay to restore post-communication tensor states, full trace is used as a selective diagnostic path rather than the default collection mode.

\section{Related Work}

Related work spans three directions: performance modeling and simulation, execution-based emulation, and distributed replay and diagnosis. RankLens connects execution-based emulation with local replay to support numerical inspection and execution analysis on a single GPU.

\noindent\textbf{Performance modeling and simulation.}
Prior work develops performance models and simulators to analyze distributed training and evaluate configuration and system design choices~\cite{guo2024survey,svedas2025survey}. Paleo, AMPeD, and Calculon use analytical models, while Habitat and NeuSight use profiling information or learned models to predict training performance~\cite{qi2017paleo,moolchandani2023amped,isaev2023calculon,geoffrey2021habitat,lee2025neusight}. Daydream, dPRO, and Lumos use trace-derived dependency graphs to analyze bottlenecks and evaluate optimizations~\cite{zhu2020daydream,hu2022dpro,liang2025lumos}. ASTRA-Sim, SimAI, Proteus, TrioSim, and Multiverse simulate distributed training workloads to study computation, communication, and system configurations~\cite{rashidi2020astra,won2023astra,wang2025simai,duan2024proteus,li2025triosim,gui2025multiverse}. RankLens focuses on executing the target rank's real local computation with recorded communication effects, making its numerical state and execution behavior directly inspectable.

\noindent\textbf{Execution-based emulation.}
Related systems reuse application or framework execution to reproduce behavior with fewer resources. SMPI executes MPI applications with simulated communication~\cite{clauss2011smpi}, while DNNEmu replaces computation with predicted operator runtimes for performance emulation~\cite{wang2022dnnemu}. TorchSim uses FakeTensor and FakeProcessGroup for runtime and memory estimation~\cite{purandare2025torchsim}. Phantora reuses unmodified ML frameworks while simulating GPU computation and communication~\cite{qin2026phantora}. NeuronaBox executes a subset of nodes while emulating their external communication environment~\cite{liu2024neuronabox}. PrismLLM executes selected ranks and uses assistant GPUs to replay virtual-rank graphs and participate in real communication~\cite{xi2026prismllm}. RankLens retains one rank's real GPU computation and restores recorded communication outputs and completion timing without communicating with peer ranks. Its replay is limited to conditions under which the recorded effects remain applicable.

\noindent\textbf{Distributed replay and diagnosis.}
MPI data replay enables independent process replay from recorded communication data~\cite{maruyama2005datareplay}. MPIWiz combines data and order recording~\cite{xue2009mpiwiz}, while ReMPI records message-matching order for deterministic replay~\cite{sato2015rempi}. Flight Recorder uses communication records to diagnose stalled jobs and collective timeouts~\cite{liu2026flightrecorder}. RankLens builds on local replay by combining logical distributed context, communication-result restoration, and asynchronous completion control in GPU training frameworks. This supports numerical inspection and analysis of computation--communication overlap and synchronization behavior within the original program.

\section{Conclusion}

We presented \textbf{RankLens}, a single-rank emulation system for debugging and tuning distributed training on a single GPU. Its central insight is that rank-local analysis can preserve the required effects of remote execution at communication boundaries while retaining real local computation. RankLens combines distributed state virtualization, restoration of recorded communication results and recorded or estimated timing, and asynchronous completion control to execute the original training program along the target rank's distributed path. This enables developers to inspect numerical anomalies, analyze computation-communication overlap and synchronization behavior, and evaluate local scheduling changes using existing debugging and performance analysis tools.

Across DeepSpeed and Megatron-LM configurations spanning 8--64 GPUs, RankLens achieves a 100\% match in training-tensor communication signature counts and keeps relative step-time error below 5\% with both recorded and estimated communication timing. Controlled experiments further demonstrate reproduction and inspection of a communicated gradient anomaly and its effect on parameter updates, as well as identification of a beneficial computation-communication scheduling change. These results show that RankLens enables repeated local debugging and tuning on a single GPU, reducing dependence on target-scale cluster executions.

\section{Acknowledgment}
The authors used ChatGPT~\cite{openai_chatgpt} solely for
English language polishing and grammatical refinement.
All technical content was reviewed and verified by the authors.

\bibliographystyle{IEEEtran}
\bibliography{references}

@IEEEtranBSTCTL{BSTcontrol,
    CTLuse_forced_etal = "yes",
    CTLmax_names_forced_etal = "10",
    CTLnames_show_etal = "3"
}

@article{grattafiori2024llama,
  author       = {Grattafiori, Aaron and Dubey, Abhimanyu and Jauhri, Abhinav and Pandey, Abhinav and Kadian, Abhishek and Al-Dahle, Ahmad and Letman, Aiesha and Mathur, Akhil and Schelten, Alan and Vaughan, Alex and Yang, Amy and Fan, Angela and Goyal, Anirudh and Hartshorn, Anthony and Yang, Aobo and Mitra, Archi and Sravankumar, Archie and Korenev, Artem and Hinsvark, Arthur and Rao, Arun and Zhang, Aston and Rodriguez, Aurelien and Gregerson, Austen and Spataru, Ava and Roziere, Baptiste and Biron, Bethany and Tang, Binh and Chern, Bobbie and Caucheteux, Charlotte and Nayak, Chaya and Bi, Chloe and Marra, Chris and McConnell, Chris and Keller, Christian and Touret, Christophe and Wu, Chunyang and Wong, Corinne and Ferrer, Cristian Canton and Nikolaidis, Cyrus and Allonsius, Damien and Song, Daniel and Pintz, Danielle and Livshits, Danny and Wyatt, Danny and Esiobu, David and Choudhary, Dhruv and Mahajan, Dhruv and Garcia-Olano, Diego and Perino, Diego and Hupkes, Dieuwke and Lakomkin, Egor and AlBadawy, Ehab and Lobanova, Elina and Dinan, Emily and Smith, Eric Michael and Radenovic, Filip and Guzm{\'{a}}n, Francisco and Zhang, Frank and Synnaeve, Gabriel and Lee, Gabrielle and Anderson, Georgia Lewis and Thattai, Govind and Nail, Graeme and Mialon, Gregoire and Pang, Guan and Cucurell, Guillem and Nguyen, Hailey and Korevaar, Hannah and Xu, Hu and Touvron, Hugo and Zarov, Iliyan and Ibarra, Imanol Arrieta and Kloumann, Isabel and Misra, Ishan and Evtimov, Ivan and Zhang, Jack and Copet, Jade and Lee, Jaewon and Geffert, Jan and Vranes, Jana and Park, Jason and Mahadeokar, Jay and Shah, Jeet and van der Linde, Jelmer and Billock, Jennifer and Hong, Jenny and Lee, Jenya and Fu, Jeremy and Chi, Jianfeng and Huang, Jianyu and Liu, Jiawen and Wang, Jie and Yu, Jiecao and Bitton, Joanna and Spisak, Joe and Park, Jongsoo and Rocca, Joseph and Johnstun, Joshua and Saxe, Joshua and Jia, Junteng and Alwala, Kalyan Vasuden and Prasad, Karthik and Upasani, Kartikeya and Plawiak, Kate and Li, Ke and Heafield, Kenneth and Stone, Kevin and El-Arini, Khalid and Iyer, Krithika and Malik, Kshitiz and Chiu, Kuenley and Bhalla, Kunal and Lakhotia, Kushal and Rantala-Yeary, Lauren and van der Maaten, Laurens and Chen, Lawrence and Tan, Liang and Jenkins, Liz and Martin, Louis and Madaan, Lovish and Malo, Lubo and Blecher, Lukas and Landzaat, Lukas and de Oliveira, Luke and Muzzi, Madeline and Pasupuleti, Mahesh and Singh, Mannat and Paluri, Manohar and Kardas, Marcin and Tsimpoukelli, Maria and Oldham, Mathew and Rita, Mathieu and Pavlova, Maya and Kambadur, Melanie and Lewis, Mike and Si, Min and Singh, Mitesh Kumar and Hassan, Mona and Goyal, Naman and Torabi, Narjes and Bashlykov, Nikolay and Bogoychev, Nikolay and Chatterji, Niladri and Zhang, Ning and Duchenne, Olivier and {\c{C}}elebi, Onur and Alrassy, Patrick and Zhang, Pengchuan and Li, Pengwei and Vasic, Petar and Weng, Peter and Bhargava, Prajjwal and Dubal, Pratik and Krishnan, Praveen and Koura, Punit Singh and Xu, Puxin and He, Qing and Dong, Qingxiao and Srinivasan, Ragavan and Ganapathy, Raj and Calderer, Ramon and Cabral, Ricardo Silveira and Stojnic, Robert and Raileanu, Roberta and Maheswari, Rohan and Girdhar, Rohit and Patel, Rohit and Sauvestre, Romain and Polidoro, Ronnie and Sumbaly, Roshan and Taylor, Ross and Silva, Ruan and Hou, Rui and Wang, Rui and Hosseini, Saghar and Chennabasappa, Sahana and Singh, Sanjay and Bell, Sean and Kim, Seohyun Sonia and Edunov, Sergey and Nie, Shaoliang and Narang, Sharan and Raparthy, Sharath and Shen, Sheng and Wan, Shengye and Bhosale, Shruti and Zhang, Shun and Vandenhende, Simon and Batra, Soumya and Whitman, Spencer and Sootla, Sten and Collot, Stephane and Gururangan, Suchin and Borodinsky, Sydney and Herman, Tamar and Fowler, Tara and Sheasha, Tarek and Georgiou, Thomas and Scialom, Thomas and Speckbacher, Tobias and Mihaylov, Todor and Xiao, Tong and Karn, Ujjwal and Goswami, Vedanuj and Gupta, Vibhor and Ramanathan, Vignesh and Kerkez, Viktor and Gonguet, Vincent and Do, Virginie and Vogeti, Vish and Albiero, V{\'{i}}tor and Petrovic, Vladan and Chu, Weiwei and Xiong, Wenhan and Fu, Wenyin and Meers, Whitney and Martinet, Xavier and Wang, Xiaodong and Wang, Xiaofang and Tan, Xiaoqing Ellen and Xia, Xide and Xie, Xinfeng and Jia, Xuchao and Wang, Xuewei and Goldschlag, Yaelle and Gaur, Yashesh and Babaei, Yasmine and Wen, Yi and Song, Yiwen and Zhang, Yuchen and Li, Yue and Mao, Yuning and Coudert, Zacharie Delpierre and Yan, Zheng and Chen, Zhengxing and Papakipos, Zoe and Singh, Aaditya and Srivastava, Aayushi and Jain, Abha and Kelsey, Adam and Shajnfeld, Adam and Gangidi, Adithya and Victoria, Adolfo and Goldstand, Ahuva and Menon, Ajay and Sharma, Ajay and Boesenberg, Alex and Baevski, Alexei and Feinstein, Allie and Kallet, Amanda and Sangani, Amit and Teo, Amos and Yunus, Anam and Lupu, Andrei and Alvarado, Andres and Caples, Andrew and Gu, Andrew and Ho, Andrew and Poulton, Andrew and Ryan, Andrew and Ramchandani, Ankit and Dong, Annie and Franco, Annie and Goyal, Anuj and Saraf, Aparajita and Chowdhury, Arkabandhu and Gabriel, Ashley and Bharambe, Ashwin and Eisenman, Assaf and Yazdan, Azadeh and James, Beau and Maurer, Ben and Leonhardi, Benjamin and Huang, Bernie and Loyd, Beth and De Paola, Beto and Paranjape, Bhargavi and Liu, Bing and Wu, Bo and Ni, Boyu and Hancock, Braden and Wasti, Bram and Spence, Brandon and Stojkovic, Brani and Gamido, Brian and Montalvo, Britt and Parker, Carl and Burton, Carly and Mejia, Catalina and Liu, Ce and Wang, Changhan and Kim, Changkyu and Zhou, Chao and Hu, Chester and Chu, Ching-Hsiang and Cai, Chris and Tindal, Chris and Feichtenhofer, Christoph and Gao, Cynthia and Civin, Damon and Beaty, Dana and Kreymer, Daniel and Li, Daniel and Adkins, David and Xu, David and Testuggine, Davide and David, Delia and Parikh, Devi and Liskovich, Diana and Foss, Didem and Wang, Dingkang and Le, Duc and Holland, Dustin and Dowling, Edward and Jamil, Eissa and Montgomery, Elaine and Presani, Eleonora and Hahn, Emily and Wood, Emily and Le, Eric-Tuan and Brinkman, Erik and Arcaute, Esteban and Dunbar, Evan and Smothers, Evan and Sun, Fei and Kreuk, Felix and Tian, Feng and Kokkinos, Filippos and Ozgenel, Firat and Caggioni, Francesco and Kanayet, Frank and Seide, Frank and Florez, Gabriela Medina and Schwarz, Gabriella and Badeer, Gada and Swee, Georgia and Halpern, Gil and Herman, Grant and Sizov, Grigory and Zhang, Guangyi (Jack) and Lakshminarayanan, Guna and Inan, Hakan and Shojanazeri, Hamid and Zou, Han and Wang, Hannah and Zha, Hanwen and Habeeb, Haroun and Rudolph, Harrison and Suk, Helen and Aspegren, Henry and Goldman, Hunter and Zhan, Hongyuan and Damlaj, Ibrahim and Molybog, Igor and Tufanov, Igor and Leontiadis, Ilias and Veliche, Irina-Elena and Gat, Itai and Weissman, Jake and Geboski, James and Kohli, James and Lam, Janice and Asher, Japhet and Gaya, Jean-Baptiste and Marcus, Jeff and Tang, Jeff and Chan, Jennifer and Zhen, Jenny and Reizenstein, Jeremy and Teboul, Jeremy and Zhong, Jessica and Jin, Jian and Yang, Jingyi and Cummings, Joe and Carvill, Jon and Shepard, Jon and McPhie, Jonathan and Torres, Jonathan and Ginsburg, Josh and Wang, Junjie and Wu, Kai and U, Kam Hou and Saxena, Karan and Khandelwal, Kartikay and Zand, Katayoun and Matosich, Kathy and Veeraraghavan, Kaushik and Michelena, Kelly and Li, Keqian and Jagadeesh, Kiran and Huang, Kun and Chawla, Kunal and Huang, Kyle and Chen, Lailin and Garg, Lakshya and A, Lavender and Silva, Leandro and Bell, Lee and Zhang, Lei and Guo, Liangpeng and Yu, Licheng and Moshkovich, Liron and Wehrstedt, Luca and Khabsa, Madian and Avalani, Manav and Bhatt, Manish and Mankus, Martynas and Hasson, Matan and Lennie, Matthew and Reso, Matthias and Groshev, Maxim and Naumov, Maxim and Lathi, Maya and Keneally, Meghan and Liu, Miao and Seltzer, Michael L. and Valko, Michal and Restrepo, Michelle and Patel, Mihir and Vyatskov, Mik and Samvelyan, Mikayel and Clark, Mike and Macey, Mike and Wang, Mike and Hermoso, Miquel Jubert and Metanat, Mo and Rastegari, Mohammad and Bansal, Munish and Santhanam, Nandhini and Parks, Natascha and White, Natasha and Bawa, Navyata and Singhal, Nayan and Egebo, Nick and Usunier, Nicolas and Mehta, Nikhil and Laptev, Nikolay Pavlovich and Dong, Ning and Cheng, Norman and Chernoguz, Oleg and Hart, Olivia and Salpekar, Omkar and Kalinli, Ozlem and Kent, Parkin and Parekh, Parth and Saab, Paul and Balaji, Pavan and Rittner, Pedro and Bontrager, Philip and Roux, Pierre and Dollar, Piotr and Zvyagina, Polina and Ratanchandani, Prashant and Yuvraj, Pritish and Liang, Qian and Alao, Rachad and Rodriguez, Rachel and Ayub, Rafi and Murthy, Raghotham and Nayani, Raghu and Mitra, Rahul and Parthasarathy, Rangaprabhu and Li, Raymond and Hogan, Rebekkah and Battey, Robin and Wang, Rocky and Howes, Russ and Rinott, Ruty and Mehta, Sachin and Siby, Sachin and Bondu, Sai Jayesh and Datta, Samyak and Chugh, Sara and Hunt, Sara and Dhillon, Sargun and Sidorov, Sasha and Pan, Satadru and Mahajan, Saurabh and Verma, Saurabh and Yamamoto, Seiji and Ramaswamy, Sharadh and Lindsay, Shaun and Lindsay, Shaun and Feng, Sheng and Lin, Shenghao and Zha, Shengxin Cindy and Patil, Shishir and Shankar, Shiva and Zhang, Shuqiang and Zhang, Shuqiang and Wang, Sinong and Agarwal, Sneha and Sajuyigbe, Soji and Chintala, Soumith and Max, Stephanie and Chen, Stephen and Kehoe, Steve and Satterfield, Steve and Govindaprasad, Sudarshan and Gupta, Sumit and Deng, Summer and Cho, Sungmin and Virk, Sunny and Subramanian, Suraj and Choudhury, Sy and Goldman, Sydney and Remez, Tal and Glaser, Tamar and Best, Tamara and Koehler, Thilo and Robinson, Thomas and Li, Tianhe and Zhang, Tianjun and Matthews, Tim and Chou, Timothy and Shaked, Tzook and Vontimitta, Varun and Ajayi, Victoria and Montanez, Victoria and Mohan, Vijai and Kumar, Vinay Satish and Mangla, Vishal and Ionescu, Vlad and Poenaru, Vlad and Mihailescu, Vlad Tiberiu and Ivanov, Vladimir and Li, Wei and Wang, Wenchen and Jiang, Wenwen and Bouaziz, Wes and Constable, Will and Tang, Xiaocheng and Wu, Xiaojian and Wang, Xiaolan and Wu, Xilun and Gao, Xinbo and Kleinman, Yaniv and Chen, Yanjun and Hu, Ye and Jia, Ye and Qi, Ye and Li, Yenda and Zhang, Yilin and Zhang, Ying and Adi, Yossi and Nam, Youngjin and Wang, Yu (Sid) and Zhao, Yu and Hao, Yuchen and Qian, Yundi and Li, Yunlu and He, Yuzi and Rait, Zach and DeVito, Zachary and Rosnbrick, Zef and Wen, Zhaoduo and Yang, Zhenyu and Zhao, Zhiwei and Ma, Zhiyu},
  title        = {The {Llama 3} Herd of Models},
  journal      = {arXiv preprint arXiv:2407.21783},
  year         = {2024},
  doi          = {10.48550/arXiv.2407.21783},
  url          = {https://doi.org/10.48550/arXiv.2407.21783},
}

@article{deepseek2026v4,
  author       = {{DeepSeek-AI} and Xu, Anyi and Lin, Bangcai and Xue, Bing and Wang, Bingxuan and Xu, Bingzheng and Wu, Bochao and Zhang, Bowei and Lin, Chaofan and Dong, Chen and Ling, Chenchen and Lu, Chengda and Zhao, Chenggang and Deng, Chengqi and Hou, Chengyu and Xu, Chenhao and Shao, Chenze and Ruan, Chong and Sun, Conner and Dai, Damai and Guo, Daya and Yang, Dejian and Chen, Deli and Li, Donghao and Ji, Dongjie and Li, Erhang and Wei, Fang and Lin, Fangyun and Yuan, Fangzhou and Xia, Feiyu and Dai, Fucong and Hao, Guangbo and Chen, Guanting and Cao, Guoai and Meng, Guolai and Li, Guowei and Yu, Han and Zhang, Han and Xu, Hanwei and Li, Hao and Liang, Haofen and Zhang, Haoling and Luo, Haoming and Wei, Haoran and Yuan, Haotian and Zhang, Haowei and Luo, Haowen and Chen, Haoyu and Ji, Haozhe and Zhang, Hengqing and Ding, Honghui and Tang, Hongxuan and Cao, Huanqi and Gao, Huazuo and Qu, Hui and Zeng, Hui and Yang, J and Zhu, JQ and Luo, Jia and Song, Jia and Yu, Jia and Huang, Jialiang and Cai, Jialu and Liang, Jian and Zhou, Jiangting and Ye, Jiasheng and Li, Jiashi and Xu, Jiaxin and Hu, Jiewen and Yang, Jieyu and Chen, Jin and Yan, Jin and Chen, Jingchang and Zhou, Jingli and Xiang, Jingting and Yuan, Jingyang and Cheng, Jingyuan and Zhou, Jingzi and Zhu, Jinhua and Yu, Jiping and Sun, Joseph and Ran, Jun and Jiang, Junguang and Qiu, Junjie and Li, Junlong and Zheng, Junmin and Song, Junxiao and Dong, Kai and Gao, Kaige and Guan, Kang and Zhou, Kexing and Huang, Kezhao and Yu, Kuai and Wang, Lean and Zhang, Lecong and Wang, Lei and Xia, Leyi and Zhang, Li and Zhao, Liang and Guo, Lihua and Luo, Lingxiao and Ma, Linwang and Zhu, Linyan and Wang, Litong and Cai, Liyu and Zhang, Liyue and Chen, Longhao and Di, MS and Xu, MY and Mei, Max and Wang, Miaojun and Zhang, Mingchuan and Zhang, Minghua and Tang, Minghui and Li, Mingming and Zhou, Mingxu and Han, Minmin and Wang, Ning and Huang, Panpan and Wang, Panpan and Cong, Peixin and Wang, Peiyi and Zhang, Peng and Wang, Qiancheng and Zhu, Qihao and Li, Qingyang and Chen, Qinyu and Du, Qiushi and Jiang, Qiwei and Tian, Rui and Xu, Ruifan and Lu, Ruijie and Xu, Ruiling and Ge, Ruiqi and Zhang, Ruisong and Pan, Ruizhe and Wang, Runji and Chen, Runqian and Yin, Runqiu and Xu, Runxin and Shen, Ruomeng and Zhang, Ruoyu and Chen, Ruyi and Liu, SH and Lu, Shanghao and Sun, Shangmian and Zhou, Shangyan and Chen, Shanhuang and Cai, Shaofei and Nie, Shaoheng and Wu, Shaoqing and Chen, Shaoyuan and Hu, Shengding and Liu, Shengyu and Hu, Shiqiang and Ma, Shirong and Wang, Shiyu and Yu, Shuiping and Zhou, Shunfeng and Pan, Shuting and Yu, Shuying and Zhou, Songyang and Ni, Tao and Yun, Tao and Jin, Tian and Pei, Tian and Ye, Tian and Lin, Tianle and Ji, Tianran and Cui, Tianyi and Yue, Tianyuan and Yu, Tingting and Wang, Tun and Zhang, W and Xiao, WL and Zeng, Wangding and An, Wei and Zhao, Weilin and Liu, Wen and Liang, Wenfeng and Pang, Wenjie and Luo, Wenjing and Yao, Wenjing and Gao, Wenjun and Yang, Wenkai and Huang, Wenlve and Hou, Wenqing and Zhang, Wentao and Ma, Wenting and Gao, Xi and He, Xiang and Wang, Xiangwen and Wang, Xianzu and Bi, Xiao and Liu, Xiaodong and Wang, Xiaohan and Chen, Xiaokang and Zhang, Xiaokang and Nie, Xiaotao and Sun, Xiaowen and Wang, Xiaoxiang and Cheng, Xin and Liu, Xin and Xie, Xin and Liu, Xingchao and Liu, Xingchen and Yu, Xingkai and Li, Xingyou and Yang, Xinyu and Zhang, Xinyu and Chen, Xu and Wang, Xuanyu and Su, Xuecheng and Chen, Xueyin and Lin, Xuheng and Fu, Xuwei and Yan, YC and Wang, YQ and Ma, YW and Luo, Yanfeng and Zhang, Yang and Xu, Yanhong and Ma, Yanru and Huang, Yanwen and Li, Yao and Li, Yao and Xu, Yao and Zhao, Yao and Sun, Yaofeng and Wang, Yaohui and Qian, Yi and Shao, Yi and Yu, Yi and Zhang, Yichao and Ding, Yifan and Shi, Yifan and Wu, Yijia and Xiong, Yiliang and Ma, Yiling and He, Ying and Tang, Ying and Zhou, Ying and Luo, Yingjia and Zhong, Yinmin and Piao, Yishi and Wang, Yisong and Zhang, Yixiang and Chen, Yixiao and Tan, Yixuan and Wei, Yixuan and Ma, Yiyang and Liu, Yiyuan and Yang, Yonglun and Guo, Yongqiang and Wu, Yongtong and Wu, Yu and Li, YuKun and Cheng, Yuan and Ou, Yuan and Xu, Yuanfan and Li, Yuanhao and Wang, Yuduan and Yang, Yuehan and Xu, Yuer and Wu, Yuhan and Meng, Yuhao and Zou, Yuheng and Zha, Yukun and Xiong, Yunfan and Chen, Yupeng and Lin, Yuping and Cao, Yuqian and Wang, Yuqian and Zhang, Yushun and Yan, Yuting and Lin, Yutong and Gu, Yuxian and Luo, Yuxiang and You, Yuxiang and Liu, Yuxuan and Zhou, Yuxuan and Zhou, Yuyang and Huang, Yuzhen and Wu, ZF and Wang, Zehao and Zhao, Zehua and Ren, Zehui and Zhang, Zekai and Sha, Zhangli and Fu, Zhe and Ju, Zhe and Xu, Zhean and Xie, Zhenda and Zhang, Zhengyan and Gao, Zheren and Hao, Zhewen and Gou, Zhibin and Ma, Zhicheng and Yan, Zhigang and Shao, Zhihong and Huang, Zhixian and Chen, Zhixuan and Wu, Zhiyu and Ren, Zhizhou and Wu, Zhongyu and Li, Zhuoshu and Zhang, Zhuping and Xu, Zian and Wang, Zihao and Qu, Zihua and Gu, Zihui and Zhu, Zijia and Li, Zilin and Zhang, Zipeng and Xie, Ziwei and Gao, Ziyi and Wan, Ziyi and Pan, Zizheng and Yao, Zongqing},
  title        = {{DeepSeek-V4}: Towards Highly Efficient Million-Token Context Intelligence},
  year         = {2026},
  url          = {https://doi.org/10.48550/arXiv.2606.19348},
  journal      = {arXiv preprint arXiv:2606.19348},
  doi          = {10.48550/arXiv.2606.19348},
}

@inproceedings{rajbhandari2020zero,
  author       = {Rajbhandari, Samyam and Rasley, Jeff and Ruwase, Olatunji and He, Yuxiong},
  title        = {{ZeRO}: Memory optimizations toward training trillion parameter models},
  booktitle    = {SC20: international conference for high performance computing, networking, storage and analysis},
  pages        = {1--16},
  year         = {2020},
  organization = {IEEE},
  doi          = {10.1109/SC41405.2020.00024},
  url          = {https://doi.org/10.1109/SC41405.2020.00024},
}

@inproceedings{narayanan2021efficient,
  author       = {Narayanan, Deepak and Shoeybi, Mohammad and Casper, Jared and LeGresley, Patrick and Patwary, Mostofa and Korthikanti, Vijay and Vainbrand, Dmitri and Kashinkunti, Prethvi and Bernauer, Julie and Catanzaro, Bryan and Phanishayee, Amar and Zaharia, Matei},
  title        = {Efficient large-scale language model training on {GPU} clusters using {Megatron-LM}},
  booktitle    = {Proceedings of the international conference for high performance computing, networking, storage and analysis},
  pages        = {1--15},
  year         = {2021},
  doi          = {10.1145/3458817.3476209},
  url          = {https://doi.org/10.1145/3458817.3476209},
}

@article{guo2024survey,
  author       = {Guo, Zhenhua and Tang, Yinan and Zhai, Jidong and Yuan, Tongtong and Jin, Jian and Wang, Li and Zhao, Yaqian and Li, Rengang},
  title        = {A survey on performance modeling and prediction for distributed {DNN} training},
  journal      = {IEEE Transactions on Parallel and Distributed Systems},
  volume       = {35},
  number       = {12},
  pages        = {2463--2478},
  year         = {2024},
  publisher    = {IEEE},
  doi          = {10.1109/TPDS.2024.3476390},
  url          = {https://doi.org/10.1109/TPDS.2024.3476390},
}

@article{svedas2025survey,
  author       = {Svedas, Jonas and Watson, Hannah and Laubeuf, Nathan and Moolchandani, Diksha and Nada, Abubakr and Singh, Arjun and Biswas, Dwaipayan and Myers, James and Bhattacharjee, Debjyoti},
  title        = {A survey of end-to-end modeling for distributed {DNN} training: Workloads, simulators, and {TCO}},
  journal      = {arXiv preprint arXiv:2506.09275},
  year         = {2025},
  doi          = {10.48550/arXiv.2506.09275},
  url          = {https://doi.org/10.48550/arXiv.2506.09275},
}

@inproceedings{moolchandani2023amped,
  author       = {Moolchandani, Diksha and Kundu, Joyjit and Ruelens, Frederik and Vrancx, Peter and Evenblij, Timon and Perumkunnil, Manu},
  title        = {{AMPeD}: An analytical model for performance in distributed training of transformers},
  booktitle    = {2023 IEEE International Symposium on Performance Analysis of Systems and Software (ISPASS)},
  pages        = {306--315},
  year         = {2023},
  organization = {IEEE},
  doi          = {10.1109/ISPASS57527.2023.00037},
  url          = {https://doi.org/10.1109/ISPASS57527.2023.00037},
}

@inproceedings{isaev2023calculon,
  author       = {Isaev, Mikhail and McDonald, Nic and Dennison, Larry and Vuduc, Richard W.},
  title        = {Calculon: a methodology and tool for high-level co-design of systems and large language models},
  booktitle    = {Proceedings of the International Conference for High Performance Computing, Networking, Storage and Analysis},
  pages        = {1--14},
  year         = {2023},
  doi          = {10.1145/3581784.3607102},
  url          = {https://doi.org/10.1145/3581784.3607102},
}

@inproceedings{qi2017paleo,
  author       = {Qi, Hang and Sparks, Evan R and Talwalkar, Ameet},
  title        = {Paleo: A performance model for deep neural networks},
  booktitle    = {International Conference on Learning Representations},
  year         = {2017},
  url          = {https://openreview.net/forum?id=SyVVJ85lg},
}

@article{hu2022dpro,
  author       = {Hu, Hanpeng and Jiang, Chenyu and Zhong, Yuchen and Peng, Yanghua and Wu, Chuan and Zhu, Yibo and Lin, Haibin and Guo, Chuanxiong},
  title        = {{dPRO}: A generic performance diagnosis and optimization toolkit for expediting distributed {DNN} training},
  journal      = {Proceedings of Machine Learning and Systems},
  volume       = {4},
  pages        = {623--637},
  year         = {2022},
  url          = {https://proceedings.mlsys.org/paper_files/paper/2022/hash/b422680f3db0986ddd7f8f126baaf0fa-Abstract.html},
}

@inproceedings{rashidi2020astra,
  author       = {Rashidi, Saeed and Sridharan, Srinivas and Srinivasan, Sudarshan and Krishna, Tushar},
  title        = {{ASTRA-Sim}: Enabling {SW/HW} co-design exploration for distributed {DL} training platforms},
  booktitle    = {2020 IEEE International Symposium on Performance Analysis of Systems and Software (ISPASS)},
  pages        = {81--92},
  year         = {2020},
  organization = {IEEE},
  doi          = {10.1109/ISPASS48437.2020.00018},
  url          = {https://doi.org/10.1109/ISPASS48437.2020.00018},
}

@inproceedings{won2023astra,
  author       = {Won, William and Heo, Taekyung and Rashidi, Saeed and Sridharan, Srinivas and Srinivasan, Sudarshan and Krishna, Tushar},
  title        = {{ASTRA-Sim2.0}: Modeling hierarchical networks and disaggregated systems for large-model training at scale},
  booktitle    = {2023 IEEE International Symposium on Performance Analysis of Systems and Software (ISPASS)},
  pages        = {283--294},
  year         = {2023},
  organization = {IEEE},
  doi          = {10.1109/ISPASS57527.2023.00035},
  url          = {https://doi.org/10.1109/ISPASS57527.2023.00035},
}

@inproceedings{wang2025simai,
  author       = {Xizheng Wang and Qingxu Li and Yichi Xu and Gang Lu and Dan Li and Li Chen and Heyang Zhou and Linkang Zheng and Sen Zhang and Yikai Zhu and Yang Liu and Pengcheng Zhang and Kun Qian and Kunling He and Jiaqi Gao and Ennan Zhai and Dennis Cai and Binzhang Fu},
  title        = {{SimAI}: Unifying Architecture Design and Performance Tuning for Large-Scale Large Language Model Training with Scalability and Precision},
  booktitle    = {22nd USENIX Symposium on Networked Systems Design and Implementation (NSDI 25)},
  pages        = {541--558},
  year         = {2025},
  url          = {https://www.usenix.org/conference/nsdi25/presentation/wang-xizheng-simai},
}

@article{duan2024proteus,
  author       = {Duan, Jiangfei and Li, Xiuhong and Xu, Ping and Zhang, Xingcheng and Yan, Shengen and Liang, Yun and Lin, Dahua},
  title        = {Proteus: Simulating the performance of distributed {DNN} training},
  journal      = {IEEE Transactions on Parallel and Distributed Systems},
  volume       = {35},
  number       = {10},
  pages        = {1867--1878},
  year         = {2024},
  publisher    = {IEEE},
  doi          = {10.1109/TPDS.2024.3443255},
  url          = {https://doi.org/10.1109/TPDS.2024.3443255},
}

@inproceedings{zhu2020daydream,
  author       = {Zhu, Hongyu and Phanishayee, Amar and Pekhimenko, Gennady},
  title        = {{Daydream}: Accurately Estimating the Efficacy of Optimizations for {DNN} Training},
  booktitle    = {2020 USENIX Annual Technical Conference (USENIX ATC 20)},
  pages        = {337--352},
  year         = {2020},
  url          = {https://www.usenix.org/conference/atc20/presentation/zhu-hongyu},
}

@inproceedings{li2025triosim,
  author       = {Li, Ying and Bao, Yuhui and Wang, Gongyu and Mei, Xinxin and Vaid, Pranav and Ghosh, Anandaroop and Jog, Adwait and Bunandar, Darius and Joshi, Ajay and Sun, Yifan},
  title        = {{TrioSim}: A lightweight simulator for large-scale {DNN} workloads on multi-{GPU} systems},
  booktitle    = {Proceedings of the 52nd Annual International Symposium on Computer Architecture},
  pages        = {1524--1538},
  year         = {2025},
  doi          = {10.1145/3695053.3731082},
  url          = {https://doi.org/10.1145/3695053.3731082},
}

@inproceedings{geoffrey2021habitat,
  author       = {Yu, Geoffrey X. and Gao, Yubo and Golikov, Pavel and Pekhimenko, Gennady},
  title        = {{Habitat}: A Runtime-Based Computational Performance Predictor for Deep Neural Network Training},
  booktitle    = {2021 USENIX Annual Technical Conference (USENIX ATC 21)},
  pages        = {503--521},
  year         = {2021},
  url          = {https://www.usenix.org/conference/atc21/presentation/yu},
}

@misc{pytorch_distributed_doc,
  author       = {{PyTorch Contributors}},
  title        = {{PyTorch Distributed Communication Package}},
  year         = {2025},
  url          = {https://docs.pytorch.org/docs/2.7/distributed.html},
  note         = {Version 2.7. Accessed: 2026-10-08},
}

@inproceedings{liang2025lumos,
  author       = {Liang, Mingyu and Kassa, Hiwot Tadese and Fu, Wenyin and Coutinho, Brian and Feng, Louis and Delimitrou, Christina},
  title        = {Lumos: Efficient Performance Modeling and Estimation for Large-scale {LLM} Training},
  booktitle    = {Proceedings of the 8th Annual Conference on Machine Learning and Systems},
  year         = {2025},
  url          = {https://proceedings.mlsys.org/paper_files/paper/2025/hash/a66caa1703fe34705a4368c3014c1966-Abstract-Conference.html},
}

@inproceedings{gui2025multiverse,
  author       = {Gui, Fei and Gao, Kaihui and Chen, Li and Li, Dan and Liu, Vincent and Zhang, Ran and Yang, Hongbing and Xiong, Dian},
  title        = {Accelerating Design Space Exploration for {LLM} Training Systems with Multi-experiment Parallel Simulation},
  booktitle    = {22nd USENIX Symposium on Networked Systems Design and Implementation (NSDI 25)},
  pages        = {473--488},
  year         = {2025},
  month        = {apr},
  publisher    = {USENIX Association},
  address      = {Philadelphia, PA},
  isbn         = {978-1-939133-46-5},
  url          = {https://www.usenix.org/conference/nsdi25/presentation/gui},
}

@inproceedings{wang2022dnnemu,
  author       = {Wang, Jian and Yu, Enda and Dong, Dezun and Pang, Zhengbin},
  title        = {{DNNEmu}: A Lightweight Performance Emulator for Distributed {DNN} Training},
  booktitle    = {Algorithms and Architectures for Parallel Processing (ICA3PP 2022)},
  pages        = {722--736},
  year         = {2023},
  publisher    = {Springer Nature Switzerland},
  doi          = {10.1007/978-3-031-22677-9_39},
  url          = {https://doi.org/10.1007/978-3-031-22677-9_39},
}

@inproceedings{lee2025neusight,
  author       = {Lee, Seonho and Phanishayee, Amar and Mahajan, Divya},
  title        = {Forecasting {GPU} Performance for Deep Learning Training and Inference},
  booktitle    = {Proceedings of the 30th ACM International Conference on Architectural Support for Programming Languages and Operating Systems},
  year         = {2025},
  doi          = {10.1145/3669940.3707265},
  url          = {https://doi.org/10.1145/3669940.3707265},
  pages        = {493--508},
}

@article{glm5_2026,
  author       = {{GLM-5-Team} and Zeng, Aohan and Lv, Xin and Hou, Zhenyu and Du, Zhengxiao and Zheng, Qinkai and Chen, Bin and Yin, Da and Ge, Chendi and Huang, Chenghua and Xie, Chengxing and Zhu, Chenzheng and Yin, Congfeng and Wang, Cunxiang and Pan, Gengzheng and Zeng, Hao and Zhang, Haoke and Wang, Haoran and Chen, Huilong and Zhang, Jiajie and Jiao, Jian and Guo, Jiaqi and Wang, Jingsen and Du, Jingzhao and Wu, Jinzhu and Wang, Kedong and Li, Lei and Fan, Lin and Zhong, Lucen and Liu, Mingdao and Zhao, Mingming and Du, Pengfan and Dong, Qian and Lu, Rui and Li, Shuang and Cao, Shulin and Liu, Song and Jiang, Ting and Chen, Xiaodong and Zhang, Xiaohan and Huang, Xuancheng and Dong, Xuezhen and Xu, Yabo and Wei, Yao and An, Yifan and Niu, Yilin and Zhu, Yitong and Wen, Yuanhao and Cen, Yukuo and Bai, Yushi and Qiao, Zhongpei and Wang, Zihan and Wang, Zikang and Zhu, Zilin and Liu, Ziqiang and Li, Zixuan and Wang, Bojie and Wen, Bosi and Huang, Can and Cai, Changpeng and Yu, Chao and Li, Chen and Hu, Chengwei and Zhang, Chenhui and Zhang, Dan and Lin, Daoyan and Yang, Dayong and Wang, Di and Ai, Ding and Zhu, Erle and Yi, Fangzhou and Chen, Feiyu and Wen, Guohong and Sun, Hailong and Zhao, Haisha and Hu, Haiyi and Zhang, Hanchen and Liu, Hanrui and Zhang, Hanyu and Peng, Hao and Tai, Hao and Zhang, Haobo and Liu, He and Wang, Hongwei and Yan, Hongxi and Ge, Hongyu and Liu, Huan and Chu, Huanpeng and Zhao, Jia'ni and Wang, Jiachen and Zhao, Jiajing and Ren, Jiamin and Wang, Jiapeng and Zhang, Jiaxin and Gui, Jiayi and Zhao, Jiayue and Li, Jijie and An, Jing and Li, Jing and Yuan, Jingwei and Du, Jinhua and Liu, Jinxin and Zhi, Junkai and Duan, Junwen and Zhou, Kaiyue and Wei, Kangjian and Wang, Ke and Luo, Keyun and Zhang, Laiqiang and Sha, Leigang and Xu, Liang and Wu, Lindong and Ding, Lintao and Chen, Lu and Li, Minghao and Lin, Nianyi and Ta, Pan and Zou, Qiang and Song, Rongjun and Yang, Ruiqi and Tu, Shangqing and Yang, Shangtong and Wu, Shaoxiang and Zhang, Shengyan and Li, Shijie and Li, Shuang and Fan, Shuyi and Qin, Wei and Tian, Wei and Zhang, Weining and Yu, Wenbo and Liang, Wenjie and Kuang, Xiang and Cheng, Xiangmeng and Li, Xiangyang and Yan, Xiaoquan and Hu, Xiaowei and Ling, Xiaoying and Fan, Xing and Xia, Xingye and Zhang, Xinyuan and Zhang, Xinze and Pan, Xirui and Zou, Xu and Zhang, Xunkai and Liu, Yadi and Wu, Yandong and Li, Yanfu and Wang, Yidong and Zhu, Yifan and Tan, Yijun and Zhou, Yilin and Pan, Yiming and Zhang, Ying and Su, Yinpei and Geng, Yipeng and Yan, Yong and Tan, Yonglin and Bi, Yuean and Shen, Yuhan and Yang, Yuhao and Li, Yujiang and Liu, Yunan and Wang, Yunqing and Li, Yuntao and Wu, Yurong and Zhang, Yutao and Duan, Yuxi and Zhang, Yuxuan and Liu, Zezhen and Jiang, Zhengtao and Yan, Zhenhe and Zhang, Zheyu and Wei, Zhixiang and Chen, Zhuo and Feng, Zhuoer and Yao, Zijun and Chai, Ziwei and Wang, Ziyuan and Zhang, Zuzhou and Xu, Bin and Huang, Minlie and Wang, Hongning and Li, Juanzi and Dong, Yuxiao and Tang, Jie},
  title        = {{GLM-5}: From Vibe Coding to Agentic Engineering},
  journal      = {arXiv preprint arXiv:2602.15763},
  year         = {2026},
  doi          = {10.48550/arXiv.2602.15763},
  url          = {https://doi.org/10.48550/arXiv.2602.15763},
}

@article{kimi_k25_2026,
  author       = {{Kimi Team} and Bai, Tongtong and Bai, Yifan and Bao, Yiping and Cai, S. H. and Cao, Yuan and Chai, Ziwei and Charles, Y. and Che, H. S. and Chen, Cheng and Chen, Guanduo and Chen, Huarong and Chen, Jia and Chen, Jianlong and Chen, Jun and Chen, Kefan and Chen, Liang and Chen, Ruijue and Chen, Xinhao and Chen, Yanru and Chen, Yanxu and Chen, Yicun and Chen, Yimin and Chen, Yingjiang and Chen, Yuankun and Chen, Yujie and Chen, Yutian and Chen, Zhirong and Chen, Ziwei and Cheng, Dazhi and Cheng, Yean and Chu, Minghan and Cui, Jialei and Deng, Jiaqi and Diao, Muxi and Ding, Hao and Dong, Mengfan and Dong, Mengnan and Dong, Yuxin and Dong, Yuhao and Du, Angang and Du, Chenzhuang and Du, Dikang and Du, Lingxiao and Du, Yulun and Fan, Yu and Fang, Shengjun and Feng, Qiulin and Feng, Yichen and Fu, Garimugai and Fu, Kelin and Gao, Hongcheng and Gao, Tong and Ge, Yuyao and Geng, Shangyi and Gong, Chengyang and Gong, Xiaochen and Gongque, Zhuoma and Gu, Qizheng and Gu, Xinran and Gu, Yicheng and Guan, Longyu and Guan, Shuhao and Guo, Yuanying and Hao, Xiaoru and He, Dailan and He, Tianhong and He, Weiran and He, Wenyang and He, Yibo and He, Yunjia and Hong, Chao and Hu, Hao and Hu, Jiaxi and Hu, Yangyang and Hu, Zhenxing and Huang, Ke and Huang, Ruiyuan and Huang, Weixiao and Huang, Zhiqi and Jia, Chaobo and Jiang, Tao and Jiang, Zhejun and Jin, Xinyi and Jing, Yu and Lai, Guokun and Li, Aidi and Li, C. and Li, Cheng and Li, Fang and Li, Guanghe and Li, Guanyu and Li, Haitao and Li, Haoyang and Li, Jia and Li, Jingwei and Li, Junxiong and Li, Lincan and Li, Mo and Li, Weihong and Li, Wentao and Li, Xinhang and Li, Xinhao and Li, Yang and Li, Yanhao and Li, Yiwei and Li, Yuxiao and Li, Zhaowei and Li, Zhaoxi and Li, Zheming and Liao, Weilong and Lin, Jiawei and Lin, Xiaohan and Lin, Yibo and Lin, Zhishan and Lin, Zichao and Liu, Cheng and Liu, Chenyu and Liu, Hongzhang and Liu, Liang and Liu, Shaowei and Liu, Shudong and Liu, Shuran and Liu, Tianwei and Liu, Tianyu and Liu, Weizhou and Liu, Xiangyan and Liu, Yangyang and Liu, Yanming and Liu, Yibo and Liu, Yuanxin and Liu, Zhengying and Liu, Zhongnuo and Lu, Enzhe and Lu, Haoyu and Lu, Zhiyuan and Luo, G. and Luo, Junyu and Luo, Tongxu and Luo, Yashuo and Ma, Long and Mao, Shaoguang and Mei, Yuan and Men, Xin and Meng, Fanqing and Meng, Zhiyong and Miao, Yibo and Ni, Minqing and Ouyang, Kun and Pan, Siyuan and Pang, Bo and Qian, Yuchao and Qin, Ruoyu and Qin, Zeyu and Qiu, Jiezhong and Qu, Bowen and Shang, Zeyu and Shao, Youbo and Shen, Tianxiao and Shen, Zhennan and Shi, Juanfeng and Shi, Lidong and Shi, Shengyuan and Song, Feifan and Song, Pengwei and Song, Tianhui and Song, Xiaoxi and Su, Hongjin and Su, Jianlin and Su, Zhaochen and Sui, Lin and Sun, Jinsong and Sun, Junyao and Sun, Tongyu and Sung, Flood and Tai, Yunpeng and Tang, Chuning and Tang, Heyi and Tang, Xiaojuan and Tang, Zhengyang and Tao, Jiawen and Teng, Shiyuan and Tian, Chaoran and Tian, Pengfei and Wang, Bowen and Wang, Chensi and Wang, Chuang and Wang, Congcong and Wang, Dingkun and Wang, Dinglu and Wang, Dongliang and Wang, Feng and Wang, Hailong and Wang, Haiming and Wang, Hao and Wang, Hengzhi and Wang, Huaqing and Wang, Hui and Wang, Jiahao and Wang, Jinhong and Wang, Jiuzheng and Wang, Kaixin and Wang, Linian and Wang, Qibin and Wang, Shengjie and Wang, Shuyi and Wang, Si and Wang, Wei and Wang, Xiaochen and Wang, Xinyuan and Wang, Yao and Wang, Yejie and Wang, Yipu and Wang, Yiqin and Wang, Yucheng and Wang, Yuzhi and Wang, Zhaoji and Wang, Zhaowei and Wang, Zhengtao and Wang, Zhexu and Wang, Zifan and Wang, Zihan and Wang, Zizhe and Wei, Chu and Wei, Ming and Wen, Chuan and Wen, Zichen and Wu, Chengjie and Wu, Haoning and Wu, Junyan and Wu, Rucong and Wu, Wenhao and Wu, Yuefeng and Wu, Yuhao and Wu, Yuxin and Wu, Zijian and Xiao, Chenjun and Xie, Jin and Xie, Xiaotong and Xie, Yuchong and Xing, Bowei and Xu, Boyu and Xu, Jianfan and Xu, Jing and Xu, Jinjing and Xu, L. H. and Xu, Lin and Xu, Suting and Xu, Weixin and Xu, Xinbo and Xu, Xinran and Xu, Yangchuan and Xu, Yichang and Xu, Yuemeng and Xu, Zelai and Xu, Ziyao and Yan, Junjie and Yan, Yuzi and Yang, Guangyao and Yang, Hao and Yang, Junwei and Yang, Kai and Yang, Ningyuan and Yang, Xiaofei and Yang, Xinlong and Yang, Xinyu and Yang, Ying and Yang, Yi and Yang, Yi and Yang, Zhen and Yang, Zhilin and Yang, Zonghan and Yao, Haotian and Ye, Dan and Ye, Haoran and Ye, Wenjie and Ye, Zhuorui and Yebo, Peng and Yin, Bohong and Yu, Chengzhen and Yu, Longhui and Yu, Tao and Yu, Tianxiang and Yuan, Enming and Yuan, Mengjie and Yuan, Xiaokun and Yue, Yang and Zeng, Weihao and Zha, Dunyuan and Zhan, Haobing and Zhang, Dehao and Zhang, Hao and Zhang, Jin and Zhang, Puqi and Zhang, Qiao and Zhang, Rui and Zhang, Xiaobin and Zhang, Xiaoyun and Zhang, Y. and Zhang, Yadong and Zhang, Yangkun and Zhang, Yichi and Zhang, Yizhi and Zhang, Yongting and Zhang, Yu and Zhang, Yushun and Zhang, Yutao and Zhang, Yutong and Zhang, Zheng and Zhao, Chenguang and Zhao, Feifan and Zhao, Jinxiang and Zhao, Shuai and Zhao, Xiangyu and Zhao, Xuanle and Zhao, Yikai and Zhao, Zijia and Zheng, Huabin and Zheng, Ruihan and Zheng, Shaojie and Zheng, Tengyang and Zhong, Junfeng and Zhong, Longguang and Zhong, Weiming and Zhou, M. and Zhou, Runjie and Zhou, Xinyu and Zhou, Zaida and Zhu, Jinguo and Zhu, Liya and Zhu, Xinhao and Zhu, Yuxuan and Zhu, Zhen and Zhuang, Jingze and Zhuang, Weiyu and Zou, Ying and Zu, Xinxing},
  title        = {{Kimi K2.5}: Visual Agentic Intelligence},
  journal      = {arXiv preprint arXiv:2602.02276},
  year         = {2026},
  doi          = {10.48550/arXiv.2602.02276},
  url          = {https://doi.org/10.48550/arXiv.2602.02276},
}

@article{li2020pytorch,
  author       = {Li, Shen and Zhao, Yanli and Varma, Rohan and Salpekar, Omkar and Noordhuis, Pieter and Li, Teng and Paszke, Adam and Smith, Jeff and Vaughan, Brian and Damania, Pritam and Chintala, Soumith},
  title        = {{PyTorch Distributed}: Experiences on Accelerating Data Parallel Training},
  journal      = {Proceedings of the VLDB Endowment},
  volume       = {13},
  number       = {12},
  pages        = {3005--3018},
  year         = {2020},
  doi          = {10.14778/3415478.3415530},
  url          = {https://doi.org/10.14778/3415478.3415530},
}

@inproceedings{purandare2025torchsim,
  author       = {Purandare, Sanket and Yang, Emma and Zhao, Andrew and Wang, Qitong and Feng, Wei and Desmaison, Alban and Gu, Andrew and Liu, Tianyu and Wright, Less and Nadathur, Gokul and Idreos, Stratos},
  title        = {{TorchSim}: High Fidelity Runtime and Memory Estimation for Distributed Training},
  booktitle    = {ICML Workshop on Efficient Systems for Foundation Models (ES-FoMo)},
  year         = {2025},
  url          = {https://openreview.net/forum?id=UehoGDWviT},
}

@inproceedings{qin2026phantora,
  author       = {Qin, Jianxing and Chen, Jingrong and Kong, Xinhao and Wu, Yongji and Yuan, Tianjun and Luo, Liang and Wang, Zhaodong and Zhang, Ying and Chen, Tingjun and Lebeck, Alvin R. and Zhuo, Danyang},
  title        = {{Phantora}: Maximizing Code Reuse in Simulation-based Machine Learning System Performance Estimation},
  booktitle    = {23rd USENIX Symposium on Networked Systems Design and Implementation (NSDI 26)},
  pages        = {1809--1825},
  year         = {2026},
  publisher    = {USENIX Association},
  url          = {https://www.usenix.org/conference/nsdi26/presentation/qin},
}

@article{xi2026prismllm,
  author       = {Xi, Shaoke and Lao, ChonLam and Jia, Boyi and Gao, Jiaqi and Zhang, Zhipeng and Cao, Jiamin and Sutioso, Brian and Xu, Erci and Yu, Minlan and Ren, Kui and Li, Yong and Qian, Zhengping and Zhai, Ennan and Zhou, Jingren},
  title        = {A Few {GPUs}, A Whole Lotta Scale: Faithful {LLM} Training Emulation with {PrismLLM}},
  journal      = {arXiv preprint arXiv:2605.15617},
  year         = {2026},
  doi          = {10.48550/arXiv.2605.15617},
  url          = {https://doi.org/10.48550/arXiv.2605.15617},
}

@misc{gdb_manual,
  author       = {{Free Software Foundation}},
  title        = {Debugging with {GDB}: The {GNU} Source-Level Debugger},
  url          = {https://sourceware.org/gdb/current/onlinedocs/gdb.html/},
  note         = {Accessed: 2026-09-28},
}

@misc{debugpy,
  author       = {{Microsoft}},
  title        = {{debugpy}: A Debugger for {Python}},
  url          = {https://github.com/microsoft/debugpy},
  note         = {Accessed: 2026-09-28},
}

@misc{nsight_systems,
  author       = {{NVIDIA}},
  title        = {{Nsight Systems} User Guide},
  url          = {https://docs.nvidia.com/nsight-systems/UserGuide/index.html},
  note         = {Accessed: 2026-09-28},
}

@misc{liu2026flightrecorder,
  author       = {Liu, Phillip and Thakore, Uttam and Wang, Junjie and Yang, Justin},
  title        = {Flight Recorder: A New Lens for Understanding {NCCL} Watchdog Timeouts},
  year         = {2026},
  month        = {mar},
  howpublished = {PyTorch Blog},
  url          = {https://pytorch.org/blog/flight-recorder-a-new-lens-for-understanding-nccl-watchdog-timeouts/},
  note         = {Published March 25, 2026. Accessed: 2026-09-29},
}

@inproceedings{liu2024neuronabox,
  author       = {Liu, Banruo and Ojewale, Mubarak Adetunji and Ding, Yuhan and Canini, Marco},
  title        = {Towards a Flexible and High-Fidelity Approach to Distributed {DNN} Training Emulation},
  booktitle    = {ACM SIGOPS Asia-Pacific Workshop on Systems (APSys)},
  year         = {2024},
  doi          = {10.1145/3678015.3680478},
  url          = {https://doi.org/10.1145/3678015.3680478},
  pages        = {88--94},
}

@inproceedings{maruyama2005datareplay,
  author       = {Maruyama, Masao and Tsumura, Tomoaki and Nakashima, Hiroshi},
  title        = {Parallel Program Debugging Based on Data-Replay},
  booktitle    = {17th IASTED International Conference on Parallel and Distributed Computing and Systems},
  pages        = {151--156},
  year         = {2005},
  url          = {https://tsumulab.org/~tsumura/paper/pdebug-pdcs05.pdf},
}

@inproceedings{xue2009mpiwiz,
  author       = {Xue, Ruini and Liu, Xuezheng and Wu, Ming and Guo, Zhenyu and Chen, Wenguang and Zheng, Weimin and Zhang, Zheng and Voelker, Geoffrey M.},
  title        = {{MPIWiz}: Subgroup Reproducible Replay of {MPI} Applications},
  booktitle    = {14th ACM SIGPLAN Symposium on Principles and Practice of Parallel Programming (PPoPP)},
  pages        = {251--260},
  year         = {2009},
  doi          = {10.1145/1504176.1504213},
  url          = {https://doi.org/10.1145/1504176.1504213},
}

@inproceedings{clauss2011smpi,
  author       = {Clauss, Pierre-Nicolas and Stillwell, Mark and Genaud, Stephane and Suter, Frederic and Casanova, Henri and Quinson, Martin},
  title        = {Single Node On-Line Simulation of {MPI} Applications with {SMPI}},
  booktitle    = {IEEE International Parallel and Distributed Processing Symposium (IPDPS)},
  pages        = {664--675},
  year         = {2011},
  doi          = {10.1109/IPDPS.2011.69},
  url          = {https://doi.org/10.1109/IPDPS.2011.69},
}

@inproceedings{sato2015rempi,
  author       = {Sato, Kento and Ahn, Dong H. and Laguna, Ignacio and Lee, Gregory L. and Schulz, Martin},
  title        = {Clock Delta Compression for Scalable Order-Replay of Non-Deterministic Parallel Applications},
  booktitle    = {International Conference for High Performance Computing, Networking, Storage and Analysis (SC)},
  year         = {2015},
  doi          = {10.1145/2807591.2807642},
  url          = {https://doi.org/10.1145/2807591.2807642},
  pages        = {1--12},
}

@misc{openai_chatgpt,
  author       = {{OpenAI}},
  title        = {{ChatGPT}},
  year         = {2026},
  howpublished = {\url{https://chatgpt.com/}},
  note         = {Accessed: October 2026}
}
\end{document}